\pdfoutput=1
\documentclass{JINST}

\newcommand{\bb}{\ensuremath{\beta\beta}}

\title{SiPMs coated with TPB : coating protocol and characterization for NEXT}
\author{
V.~\'Alvarez$^{a}$, 
J.~Agramunt$^{a}$,
M.~Ball$^{a}$, 
M.~Batall\'e$^{b}$, 
J.~Bayarri$^{a}$, 
F.I.G.~Borges$^{c}$, 
H.~Bolink$^{p}$, 
H.~Brine$^{p}$, 
S.~C\'arcel$^{a}$, 
J.M.~Carmona$^{d}$, 
J.~Castel$^{d}$, 
J.M.~Catal\'a$^{e}$, 
S.~Cebri\'an$^{d}$, 
A.~Cervera$^{a}$, 
D.~Chan$^{f}$, 
C.A.N.~Conde$^{c}$, 
T.~Dafni$^{d}$, 
T.H.V.T.~Dias$^{c}$, 
J.~D\'iaz$^{a}$, 
R.~Esteve$^{e}$, 
P.~Evtoukhovitch$^{h}$, 
J.~Ferrando$^{p}$,
L.M.P.~Fernandes$^{c}$, 
P.~Ferrario$^{a}$, 
A.L.~Ferreira$^{g}$, 
E.~Ferrer-Ribas$^{i}$, 
E.D.C.~Freitas$^{c}$, 
S.A.~Garc\'ia$^{p}$, 
A.~Gil$^{a}$, 
I.~Giomataris$^{i}$, 
A.~Goldschmidt$^{f}$, 
E.~G\'omez$^{j}$, 
H.~G\'omez$^{d}$, 
J.J.~G\'omez-Cadenas$^{a}$, 
K.~Gonz\'alez$^{a}$, 
R.M.~Guti\'errez$^{j}$,
J.~Hauptman$^{k}$,
J.A.~Hernando-Morata$^{l}$, 
D.C.~Herrera$^{d}$, 
V.~Herrero$^{e}$, 
F.J.~Iguaz$^{i}$, 
I.G.~Irastorza$^{d}$, 
V.~Kalinnikov$^{h}$, 
L.~Labarga$^{m}$, 
I.~Liubarsky$^{a}$, 
J.A.M.~Lopes$^{c}$, 
D.~Lorca$^{a}$, 
M.~Losada$^{j}$, 
G.~Luzón$^{d}$, 
A.~Mar\'i$^{e}$, 
J.~Martin-Albo$^{a}$, 
A.M.~M\'endez$^{e}$, 
T.~Miller$^{f}$, 
A. Moisenko$^{h}$, 
F.~Monrabal$^{a}$, 
C.M.B.~Monteiro$^{c}$, 
J.M.~Monz\'o$^{e}$, 
F.J.~Mora$^{e}$, 
J.~Mu\~noz Vidal$^{a}$, 
H.~Natal da Luz$^{c}$, 
G.~Navarro$^{j}$, 
M.~Nebot$^{a}$, 
D.~Nygren$^{f}$, 
C.A.B.~Oliveira$^{g}$, 
R.~Palma$^{n}$, 
J.L.~P\'erez Aparicio$^{n}$,
J. P\'erez$^{a}$, 
E. Radicioni$^{q}$,
M. Quinto$^{q}$,  
J.~Renner$^{f}$, 
L.~Ripoll$^{b}$, 
A.~Rodriguez$^{d}$, 
J.~Rodriguez$^{a}$, 
F.P.~Santos$^{c}$, 
J.M.F.~dos Santos$^{c}$, 
L.~Segu\'i$^{d}$, 
L.~Serra$^{a}$, 
D.~Shuman$^{f}$, 
C.~Sofka$^{o}$, 
M.~Sorel$^{a}$, 
A.~Soriano$^{p}$, 
H.~Spieler$^{f}$, 
J.F.~Toledo$^{e}$, 
J.~Torrent Collell$^{b}$, 
A.~Tom\'as$^{d}$, 
Z.~Tsamalaidze$^{h}$, 
D.~V\'azquez$^{l}$, 
E.~Velicheva$^{h}$, 
J.F.C.A.~Veloso$^{g}$, 
J.A.~Villar$^{d}$, 
R.~Webb$^{o}$, 
T.~Weber$^{f}$, 
J.T.~White$^{o}$, 
N.~Yahlali$^{a}$\thanks{Corresponding author.}\\
\llap{$^{a}$}Instituto de Fisica Corpuscular (IFIC), CSIC \& Universidad  de Valencia,\\ 
Calle Catedr\'atico Jos\'e Beltr\'an, 2, 46980 Valencia, Spain\\
\llap{$^{b}$}Escola Polit\`ecnica Superior, Universitat de Girona,\\ 
Av.~Montilivi s/n, 17071 Girona, Spain\\
\llap{$^{c}$}Departamento de Fisica, Universidade de Coimbra,\\ 
Rua Larga, 3004-516 Coimbra, Portugal\\
\llap{$^{d}$}Lab.\ de F\'isica Nuclear y Astropart\'iculas, Universidad de Zaragoza,\\ 
Calle Pedro Cerbuna s/n, 50009 Zaragoza, Spain\\
\llap{$^{e}$}Instituto de Instrumentaci\'on para Imagen Molecular (I3M), U.\ Polit\'ecnica de Valencia,\\ 
Camino de Vera s/n, Edificio 8B, 46022 Valencia, Spain\\
\llap{$^{f}$}Lawrence Berkeley National Laboratory (LBNL)\\ 
1 Cyclotron Road, Berkeley, CA 94720, USA\\
\llap{$^{g}$} Institute of Nanostructures, Nanomodelling and Nanofabrication (i3N), Universidade de Aveiro\\
Campus de Santiago, 3810-193 Aveiro, Portugal\\
\llap{$^{h}$}Joint Institute for Nuclear Research (JINR)\\
Joliot-Curie 6, 141980 Dubna, Russia\\
\llap{$^{i}$}IRFU, Centre d'\'Etudes de Saclay (CEA Saclay),\\ 
Gif-sur-Yvette, France\\
\llap{$^{j}$}Universidad Antonio Nariño,\\
Bogot\'a, Colombia\\
\llap{$^{k}$}Iowa State University,\\
Ames, IA 50011 USA\\
\llap{$^{l}$}Universidade de Santiago de Compostela,\\
Santiago de Compostela, Spain\\
\llap{$^{m}$}Universidad Aut\'onoma de Madrid,\\
Cantoblanco, Madrid, Spain\\
\llap{$^{n}$}Universidad Polit\'ecnica de Valencia,\\
Valencia, Spain\\
\llap{$^{o}$}Department of Physics and Astronomy, Texas A\&M University,\\
College Station, TX 77843-4242TX, USA\\
\llap{$^p$} Instituto de Ciencia Molecular (ICMOL) \\
Catedrático José Beltran n$\circ 2$, 46980 Valencia, Spain.\\
\llap{$^q$} Instituto Nazionale di Física Nucleare (INFN), Sezione di Bari \\
Via E. Orabona 4, 70125 Bari, Italy.\\

 E-mail:  \email{nadia.yahlali@ific.uv.es}

}

\abstract{
Silicon photomultipliers (SiPM) are the photon detectors chosen for the tracking readout in NEXT, a neutrinoless  {\bb} decay experiment which uses a high pressure gaseous xenon time projection chamber (TPC).  The reconstruction of event track and topology in this gaseous detector is a key handle for background rejection. Among the commercially available sensors that can be used for tracking, SiPMs offer important advantages, mainly high gain, ruggedness, cost-effectiveness and radio-purity. Their main drawback, however, is their non sensitivity in the emission spectrum of the xenon scintillation (peak at 175~nm). This is overcome by coating these sensors with the organic wavelength shifter tetraphenyl butadienne (TPB). In this paper we describe the protocol developed for coating the SiPMs with TPB and the measurements performed for characterizing the coatings as well as the performance of the coated sensors in the UV-VUV range.   }


\begin{document}
%\linenumbers

\section{Introduction}
\label{intro}
The NEXT experiment aims at discovering the neutrinoless {\bb} decay of the  $^{136}$Xe isotope using a 100 kg high pressure xenon gas time projection chamber (TPC), with 90\% isotopic enrichment and operated in electroluminescence (EL) mode~\cite{CDR}. In the xenon gas TPC, the energy of the events can be recorded with a resolution better than 1\% FWHM
using low noise photomultipliers (PMTs) as photosensors~\cite{LBNL}.  Unlike liquid xenon, gaseous xenon allows in addition the possibility to record the track and topology of the two electrons emitted in the decay of $^{136}$Xe, because these tracks are about 30~cm long at 10 bar pressure. These electrons have in addition an energy deposition pattern~\cite{CDR} which makes them distinguishable from single-electron events whose energy sum up to Q$_{\bb}$ (2.458 MeV for $^{136}$Xe). This event topological signature can be recorded using small solid-state photosensors. 

\begin{figure}[tbhp!]
\begin{center}
\includegraphics[width=0.75\textwidth]{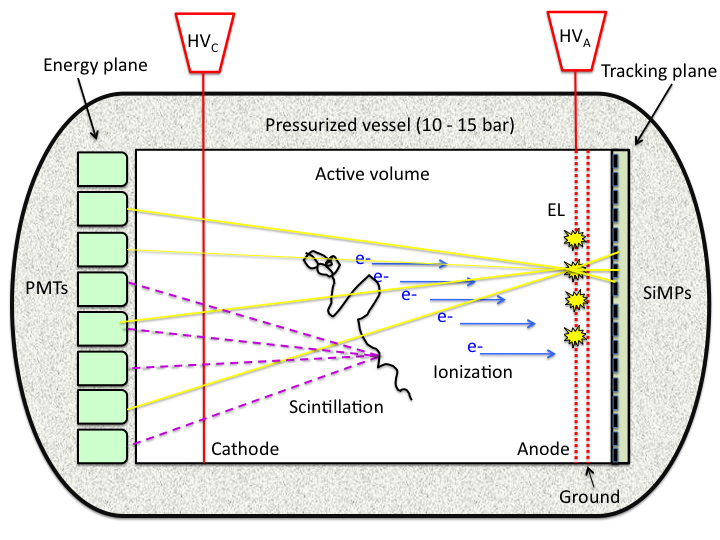} 
\end{center}
\vspace{-0.5cm}
\caption{\small The Separated Optimized Functions (SOFT) concept in NEXT TPC. EL light generated at the anode is recorded in the photosensor plane right behind it and used for tracking. It is also recorded in the photosensor plane behind the transparent cathode and used for a precise energy measurement.}
\label{fig:NEXT_tpc} 
\end{figure}

The tracking and energy measurements in NEXT are carried out by different sensors, located respectively at the TPC cathode and anode~\cite{CDR} as illustrated in figure~\ref{fig:NEXT_tpc}. 
In this figure, an event, represented as a tortuous track,  generates primary scintillation in the gaseous xenon, which is recorded primarily by 
the array of PMTs located at the TPC cathode.  
It also produces ionization electrons which drift to the TPC anode and generate EL light (or secondary scintillation), when entering the region of intense field (E/P~$\approx$~3~kV/cm.bar) between the transparent EL grids.
This light is recorded by an array of silicon photomultipliers (SiPM) located right behind the EL grids and used for tracking measurement. It is also recorded in the PMT plane behind the cathode for energy measurement. The primary scintillation recorded by PMTs gives the start-of-event time {\it t}$_0$. The EL scintillation recorded by SiPMs, provides the transversal coordinates ($x,y$) of the track's trajectory and the longitudinal coordinate ($z$) from the time {\it t} of the signal.
%%%%%
%%%%%%%
%%%%

Several NEXT prototypes with up to 1~kg of pure gaseous xenon at 10-15 bar, were recently built.  
In the NEXT-DBDM  prototype~\cite{LBNL}, the energy of the events from EL signals was measured with a near 1\% FWHM resolution from the 662 keV gamma rays of $^{137}$Cs, using an array of UV sensitive PMTs. The SiPM tracking plane first developed for the NEXT-DEMO prototype~\cite{CDR},\cite{Yahlali}, will allow to reconstruct the tracks of these gamma ray events and demonstrate that a large-mass gaseous xenon TPC, enriched with 
$^{136}$Xe and EL readout, would provide a possible pathway for a robust double-beta decay experiment.

SiPMs or Multi Pixel Photon Counters (MPPC) have been chosen in NEXT for their many outstanding features for tracking purposes. SiPMs offer comparable detection capabilities as standard small PMTs and APDs with the additional advantages of 
ruggedness, radio-purity and cost-effectiveness, essential for a large-scale radiopure detector. Their main drawback however is their poor sensitivity in the emission range of the xenon scintillation (peak at 175~nm, see reference~\cite{Gehman}). This makes necessary the use of a wavelength-shifter (WLS) to convert the UV light into visible light, where these sensors have their optimal photon detection efficiency (PDE).
 %%
%%%
\begin{figure}[h]
\centering
\includegraphics[width= 0.6\textwidth]{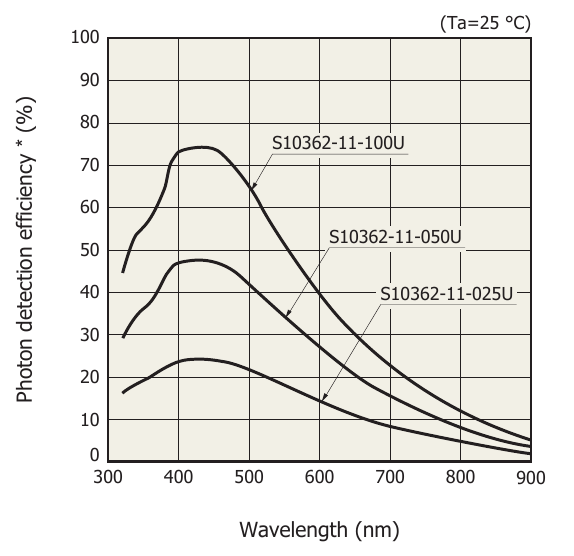}	
\caption{Photon detection efficiency as a function of wavelength of various Hamamatsu MPPCs (from \cite{Hamamatsu}).}
\label{fig:PDE}
\end{figure}
%%% 
%%%

 Tetraphenyl butadiene (TPB) is an organic compound which fluoresces when excited by UV radiation.  As reported in the literature~\cite{Lally},\cite{Gehman}, its fluorescence spectrum is peaked at about  430~nm, which matches the PDE spectrum of the SiPMs (see figure~\ref{fig:PDE}) and does not vary with the wavelength of the incident light in the UV range.
 TPB is widely used in various experiments~\cite{Boccone},\cite{Benetti} to shift scintillation light produced in the extreme UV spectrum by liquid argon or liquid xenon to the visible spectrum, where it can be detected by commercial PMTs. However, it has not been applied up to now to small solid-state photosensors. 
  
 In this paper, we describe the use of TPB coating on SiPMs to enhance their response to UV light and describe in detail the coating protocol developed for the NEXT tracking system. 
 In this system, the SiPMs are selected and calibrated to have a minimal gain dispersion ($<4\%$) over the tracking area. For this purpose, one important issue of using a WLS is to obtain sufficiently high quality coatings to conserve the uniformity of the SiPM response. The coating technique described here is particularly dedicated to the search of the conditions that guarantee uniformity, reproducibility and long-term stability of the TPB coatings on SiPMs. The coating thickness that provides the optimal fluorescence efficiency is also investigated and applied to the first prototype of the NEXT tracking system.
 In the two following sections, we describe the SiPM tracking system constructed for the NEXT-DEMO TPC prototype and the TPB-coating procedure used. In sections \ref{characterization} and \ref{response_sipms} the characterization of the TPB depositions and the response of the coated SiPMs are respectively addressed. 

%%%%%%%%%%%%%%%%%%%%%%%%%%%%    section 2   %%%%%%%%%%%%%%%%%%%%%%%%%%%%%%%%%%%%%%%%%%
%%%%%%%%%%%%%%%%%%%%%%%%%%%%                       %%%%%%%%%%%%%%%%%%%%%%%%%%%%%%%%%%%%%%%%%%
\section{NEXT tracking system} 
\label{tracking_system}
\begin{figure}[h]
\centering
\includegraphics[width= 7cm]{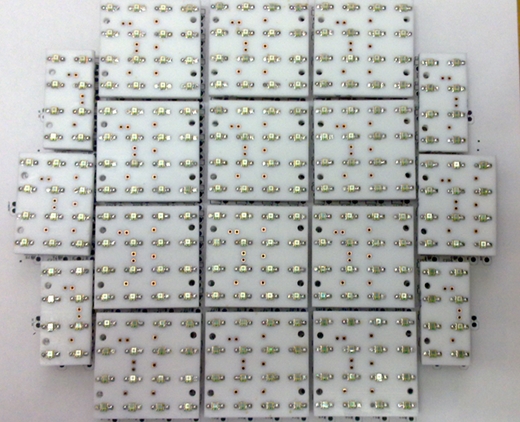}	
\includegraphics[width= 6cm]{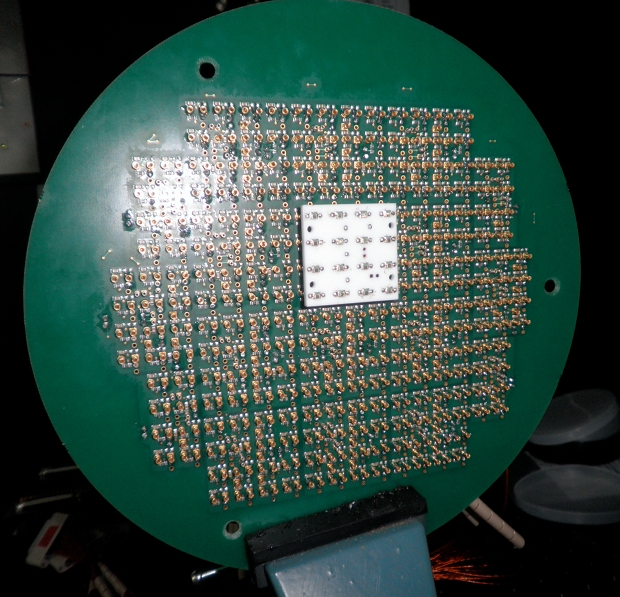}	
\caption{(left)~  SiPM daughter-boards, each containing $4\times4$ SiPMs in the center of the tracking plane and $2\times4$ or $3\times4$ SiPMs at the external edges. (right) mother-board with the bias circuits on which one of the daughter-boards is plugged.}
\label{fig:Tracking_plane}
\end{figure}

A first tracking system for NEXT was built for the TPC prototype NEXT-DEMO \cite{Yahlali}.  It is composed of 248 Hamamatsu S10362-11-025P  MPPCs \cite{Hamamatsu}, arranged to cover a circular plane of 160 mm diameter. The MPPCs are soldered onto daughter-boards (DBs) of 38$\times$38 mm$^2$ (see figure~\ref{fig:Tracking_plane}-left), made of Cuflon from Polyflon Company \cite{Polyflon}. These Cuflon boards are made of  PTFE of 3.18 mm thickness electroplated with 35~$\mu$m of oxygen-free hard copper which make them light reflective and low degassing. 
The DBs are plugged onto a large mother-board (MB) containing the printed circuits (see figure~\ref{fig:Tracking_plane}-right) for SiPM biasing. A common bias is shared by the sensors of a same DB, chosen to have very close gains with a dispersion better than 4\% at the nominal bias 
\cite{Alvarez},\cite{Herrero}.   
%%%
%%% 
The outstanding features of the chosen MPPCs are a small size (1~mm$^2$ active area), high gain ($2.75\times10^5$ at the nominal voltage $\approx 71$~V), a PDE comparable to the quantum efficiency of PMTs (25\% at 440 nm) and a wide dynamic range, due its high number of pixels (1600). The maximum number of photoelectrons (p.e.) induced by EL light expected in the SiPMs is about  250 p.e./$\mu$s (from double-beta decay electrons)~\cite{Alvarez}, which is well within their linearity range. 

The active area of the MPPCs is covered with a protective layer, usually made of a resin or PVC, which absorbs short wavelength photons, preventing them from reaching the silicon region where they can be detected \cite{Renker}. New MPPC prototypes without this protective layer are being studied in our laboratory and will be discussed in a forthcoming publication. The alternative considered here is the use of the
organic WLS fluor  1,1,4,4-Tetraphenyl-1,3-butadiene (TPB) of $\ge99$\%  purity grade \cite{Aldrich}.
This fluor in crystalline form, can be applied by vacuum-evaporation \cite{Burton} directly onto surfaces such as vessel walls or PMT windows. TPB is reported to be hard and durable, with good adherence to large substrates \cite{Burton}, \cite{Boccone}. In the following section we describe the procedure we used for coating the SiPMs with a thorough control of the coating quality.

%%%%%%%%%%%%%%%%%%%%%%%%%%%       section 3     %%%%%%%%%%%%%%%%%%%%%%%%%%%%%%%%%%%%%%%%%%%%%%%%%%%%%%%%%%%%%%%%%%%%                            %%%%%%%%%%%%%%%%%%%%%%%%%%%%%%%%%%%%%%%%%
\section{TPB coating protocol} 
\label{protocol}

\subsection{The coating system}
\begin{figure}[h]
\centering
\includegraphics[width=0.8\textwidth]{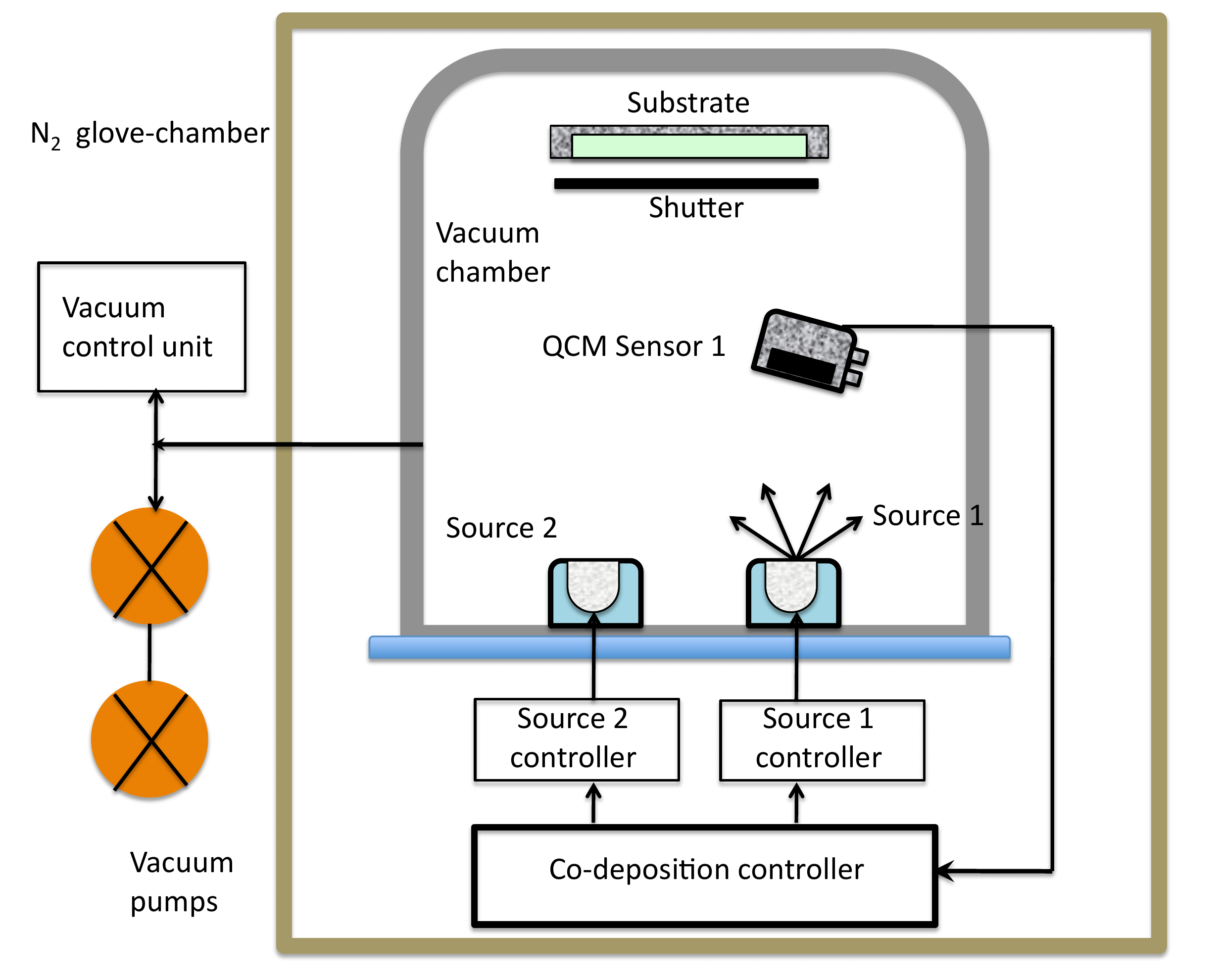}		
\caption{Scheme of the ICMOL coating system. During the TPB evaporation campaign only one source (crucible) and one QCM sensor were used. }
\label{fig:coating_system}
\end{figure}

\begin{figure}[h]
\centering	
\includegraphics[width= 0.5\textwidth]{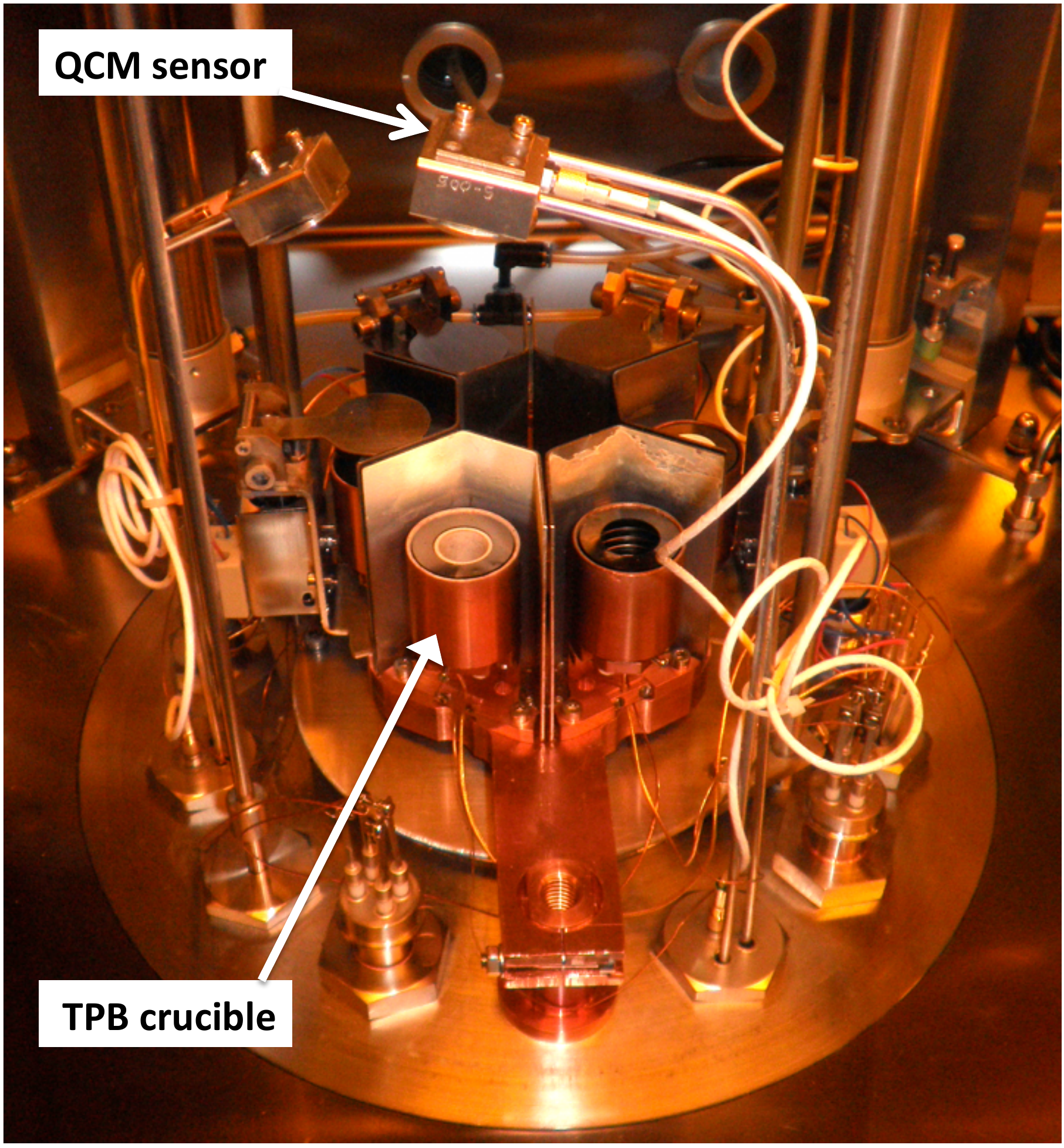}		
\caption{Picture of the evaporation system where one can distinguish the crucible used for TPB and the QCM sensor positioned on top of  it, half-way before the substrate.}
\label{fig:evaporator}
\end{figure}

The coating facility of the Instituto de Ciencia Molecular (ICMOL) was used. This facility is located in a class 10.000 clean-room due to the stringent cleanliness conditions that are required for high quality depositions of molecules on different substrates. These cleanliness conditions are particularly important in low-background applications like neutrinoless double-beta decay experiments.
In figure~\ref{fig:coating_system}, a scheme of the coating system is shown. The coating setup consists of a vacuum chamber or evaporator enclosing 4 ceramic crucibles which may melt simultaneously up to four different compounds (see figure~\ref{fig:evaporator}). The vacuum system is composed of a diaphragm pump and a turbo-molecular pump that allow vacuum levels close to $10^{-7}$~mbar in the evaporator.
The latter is enclosed in a  glove-chamber filled with N$_2$, where the manipulation of different compounds takes place in an oxygen and water-free environment to prevent oxidation and hydration. 

During the TPB evaporation campaign for NEXT, the whole setup was vacuum-cleaned to remove any traces of other molecules and only one crucible filled with TPB powder was used. The crucible was heated by a cartridge with an adjustable current that allows to monitor the temperature
and control the evaporation rate, essential to prevent bubbling and sputtering of the TPB on the substrate. 

\begin{figure}[h]
\centering
\includegraphics[width= 6.cm]{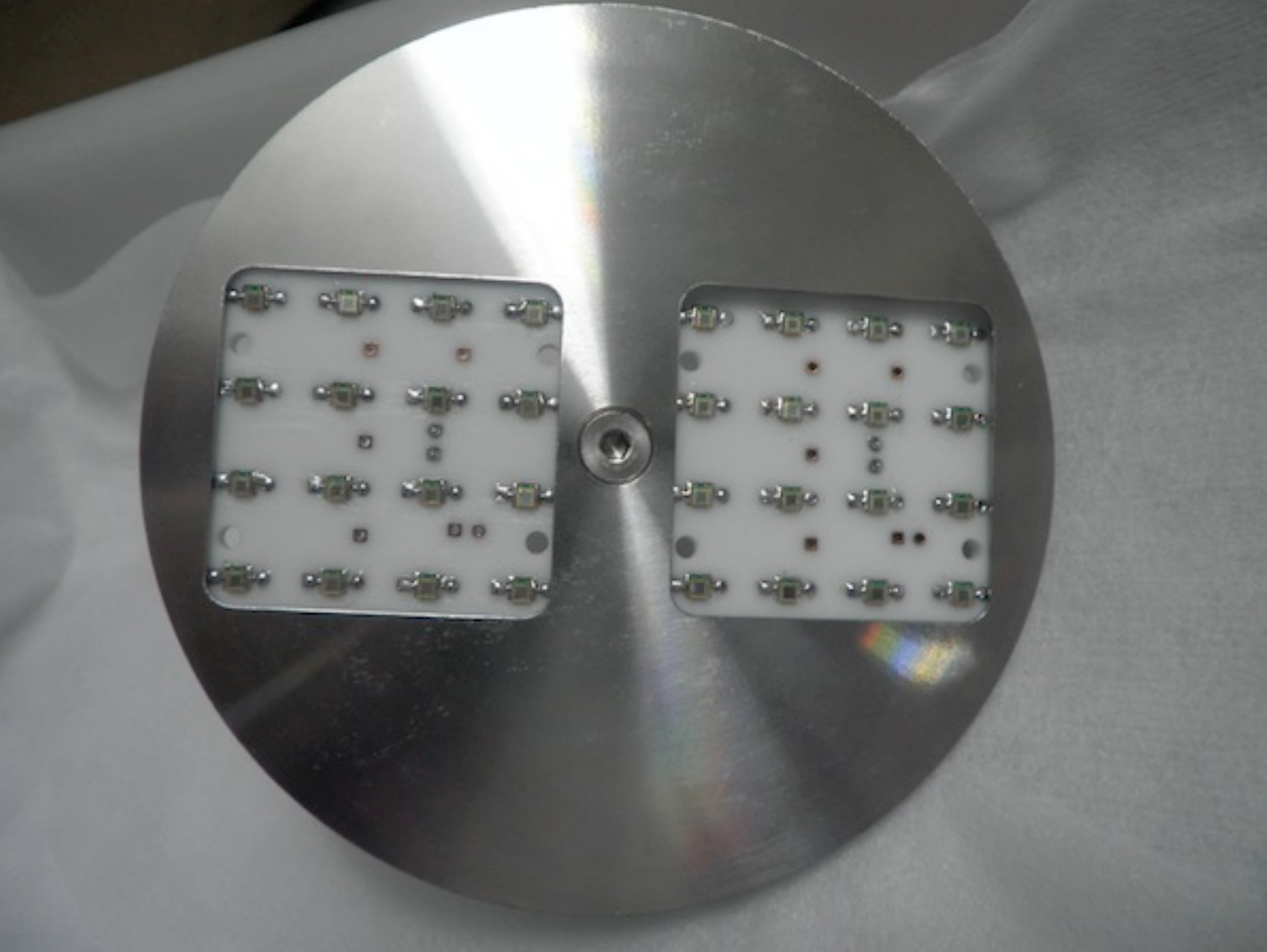}			
\caption{Picture of the sample-holder in which two SiPM DBs are lodged for coating. The holder is fixed to the spinning disk of the evaporator.} 
\label{fig:DB_support}
\end{figure}

The substrate was positioned on a sample-holder (figure~\ref{fig:DB_support}) fixed on a spinning disk located 15 cm above the crucible. A shutter located under the holder allowed to mask the exposed surface when required. After positioning the substrate, the vacuum-chamber was closed and evacuation was started.  When the optimal vacuum level is reached, typically  $4\times10^{-7}$~mbar, heating of the crucible was started with the shutter closed.
The TPB melting temperature is 203$^{\circ}$C at atmospheric pressure. At the high vacuum level reached in the evaporator, TPB evaporates at about 75$^{\circ}$C. 

The TPB deposition rate and thickness (areal mass) on the substrate were measured with ultra high precision with a Quartz Crystal Microbalance (QCM) from Sigma Instruments \cite{QCM},  located half way between the crucible and the substrate. This is a very sensitive mass deposition sensor based on the piezoelectric properties of the quartz crystal. 
The QCM is able to measure in real time mass changes ranging from micrograms to fractions of nanogram (that is a fraction of a monolayer of atoms)
on the surface of the quartz crystal.  The calibration of this sensor is thus necessary to determine accurately the deposition rate and thickness on the substrate. 

The relevant coating parameters, the deposition rate and thickness, the temperature in the crucible and the vacuum level are displayed in the deposition control units. This allowed a constant monitoring of the evaporation process.
When the deposition rate stabilized around a constant value, typically  between 1.8 and 2.4~\AA/s, a steady  evaporation process of the TPB was established.  The shutter was then opened and the spinning of the sample-holder was initiated to insure a uniform TPB deposition on the substrate. 
 When the desired thickness was reached, the shutter was closed, evacuation was stopped and the evaporator was opened. The coated samples 
 were stored in N$_2$ atmosphere or in vacuum to avoid their exposure to degrading agents.

%%%
%%% 
%%%
%%%
%%%
%%% 
%%% 

\subsection{Preparation of the substrates}

The substrates used for calibration and characterization purposes were glass plates of 30$\times$30~mm$^2$ coated with TiO$_2$ and a set of SiPM boards consisting of 5 SiPMs (Hamamatsu S10362-11-025P and -050P) soldered onto FR-4 Laminate PCBs.  These boards were used for the first TPB depositions without undergoing any special cleaning protocol.

After the characterization of these first coated samples, the SiPM daughter boards of the NEXT-DEMO tracking plane were prepared for the TPB deposition. They were first cleaned with isopropanol in an ultrasonic bath to remove dust and residues of soldering.  They were then dried in an oven at 70$^{\circ}$C during 2 hours and stored in the N$_2$ atmosphere of the glove-chamber until their introduction into the evaporator. This cleaning protocol, made in agreement with Hamamatsu, has shown to be safe for the SiPMs as no alteration of their performance (dark rate and response to a given illumination) was observed after the cleaning process. 

\subsection{Calibration}
The quartz crystal of the QCM sensor is sandwiched between two gold electrodes that are vapor deposited on either side of the crystal.  
When an alternating electric field is applied over the electrodes, the quartz crystal starts to oscillate (5 to 6 MHz in the QCM used here).  As mass is deposited on the surface of the crystal (during vacuum evaporation of TPB for instance), the oscillation frequency decreases proportionally from the initial value. 
The QCM technique, developed by  G\"unter Sauerbrey \cite{Sauerbrey}, uses this frequency shift to measure the deposition mass on the quartz crystal.
Thus the precise correlation between the deposition rate on the quartz crystal and on the substrate has to be determined.

During the TPB coating campaign, this calibration was performed using a high resolution surface profilometer (XP-1 from Ambios Technology \cite{Ambios}). 
The profile of a TPB deposition on a glass substrate, scratched with a cutter, was recorded by the profilometer. This allowed to measure the thickness of
the TPB deposition in the \AA~range. 
This thickness measurement, compared to that recorded by the QCM sensor in the evaporator, provided the calibration factor for the rate and areal mass deposition of TPB on the substrates. 

\begin{figure}[h]
\centering
\includegraphics[width= 6.5cm]{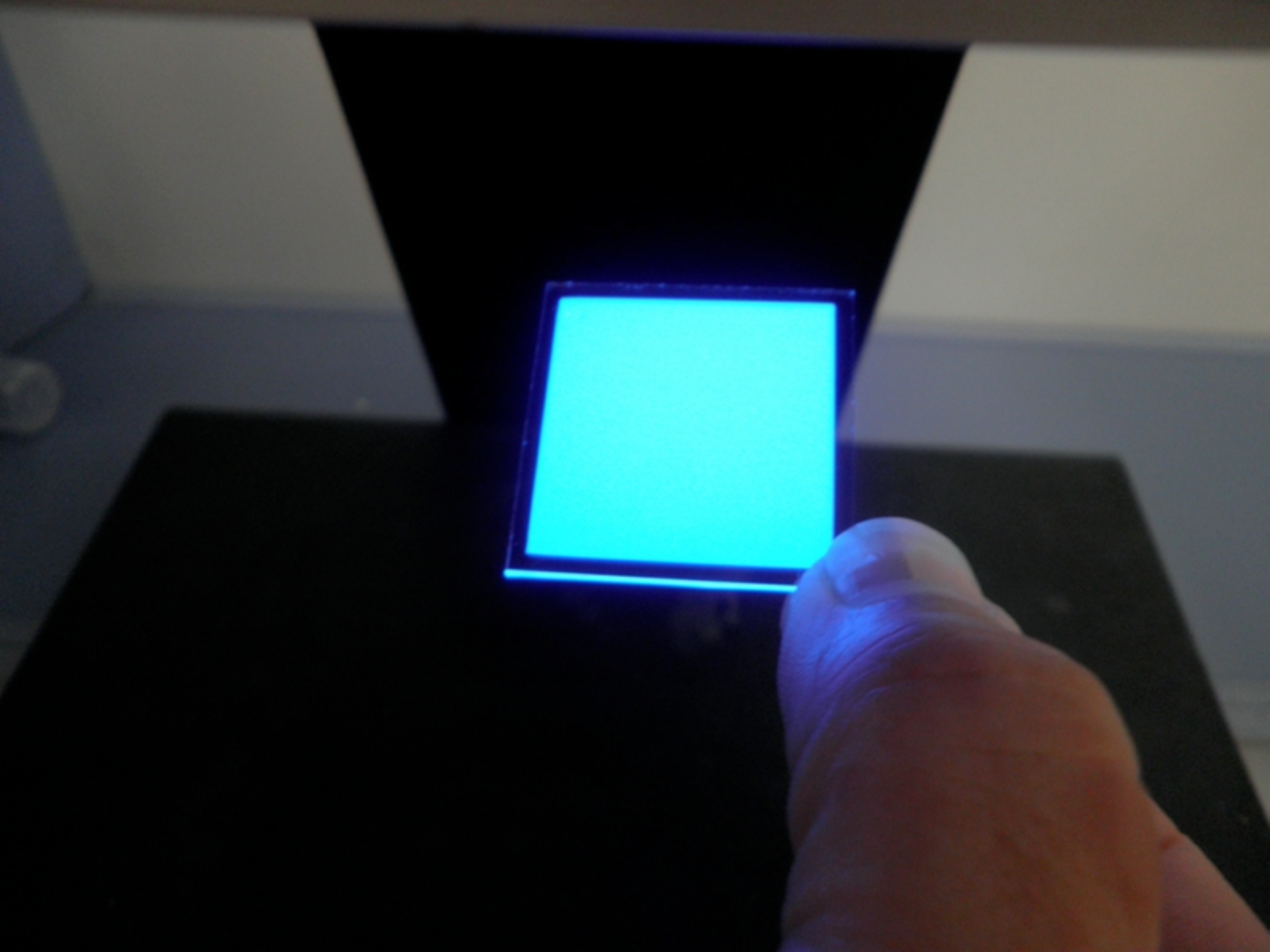}	
\includegraphics[width= 6.cm]{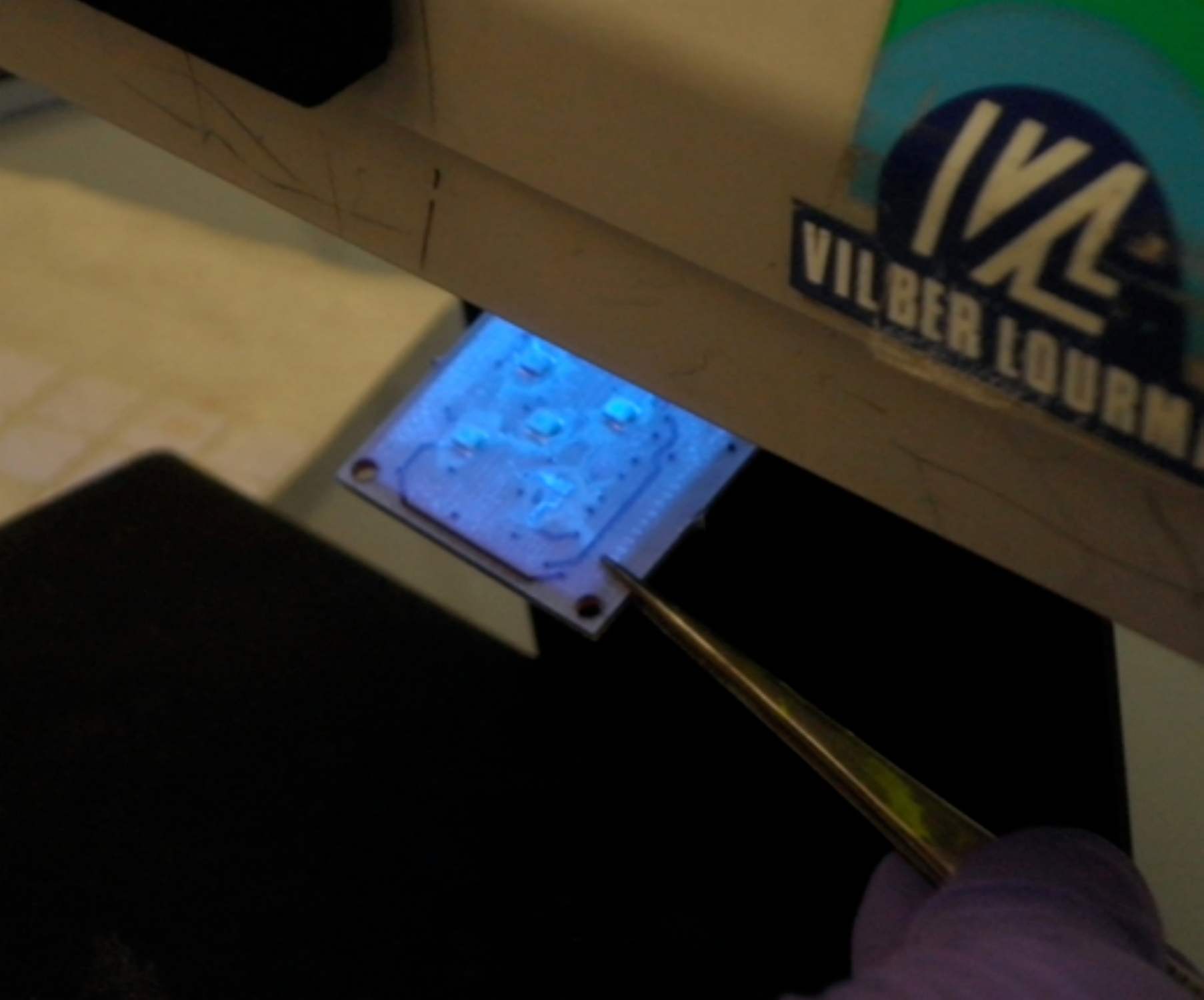}		
\caption{ A glass plate (left) and a 5-SiPMs board (right), both coated with TPB, emitting in the blue when illuminated with UV light at 240~nm. The different coating thickness and surface quality of these two samples result in a different emission yield captured by the photographs.}
\label{fig:UV_illumination}
\end{figure}

Several depositions of a chosen thickness were successfully produced on glass plates and on 5-SiPM boards.  
The coating quality showed to be reproducible as long as the batch of TPB powder used was stored in appropriate environmental conditions (see section~\ref{sec:ageing}). 
In figure~\ref{fig:UV_illumination}, a glass plate (left) and a 5-SiPM board (right) coated with TPB and illuminated with UV light at 240~nm,  clearly appear re-emitting in the blue.
The substrate samples coated with different TPB thicknesses were tested and characterized with different UV light sources. 
%%%
%%% 

%%%%%%%%%%%%%%%%%%%%%%%%%%%%%%       section 4     %%%%%%%%%%%%%%%%%%%%%%%%%%%%%%%%%%%%%%%%%%%%%%%%%%%%%%%%%%%%%%%%%%%%                            %%%%%%%%%%%%%%%%%%%%%%%%%%%%%%%%%%%%%%
\section{Characterization of TPB coatings}
\label{characterization}

\subsection{TPB emission spectrum} 
\label{sec:fluorescence}

\begin{figure}[h]
\centering
\includegraphics[width= 0.75\textwidth]{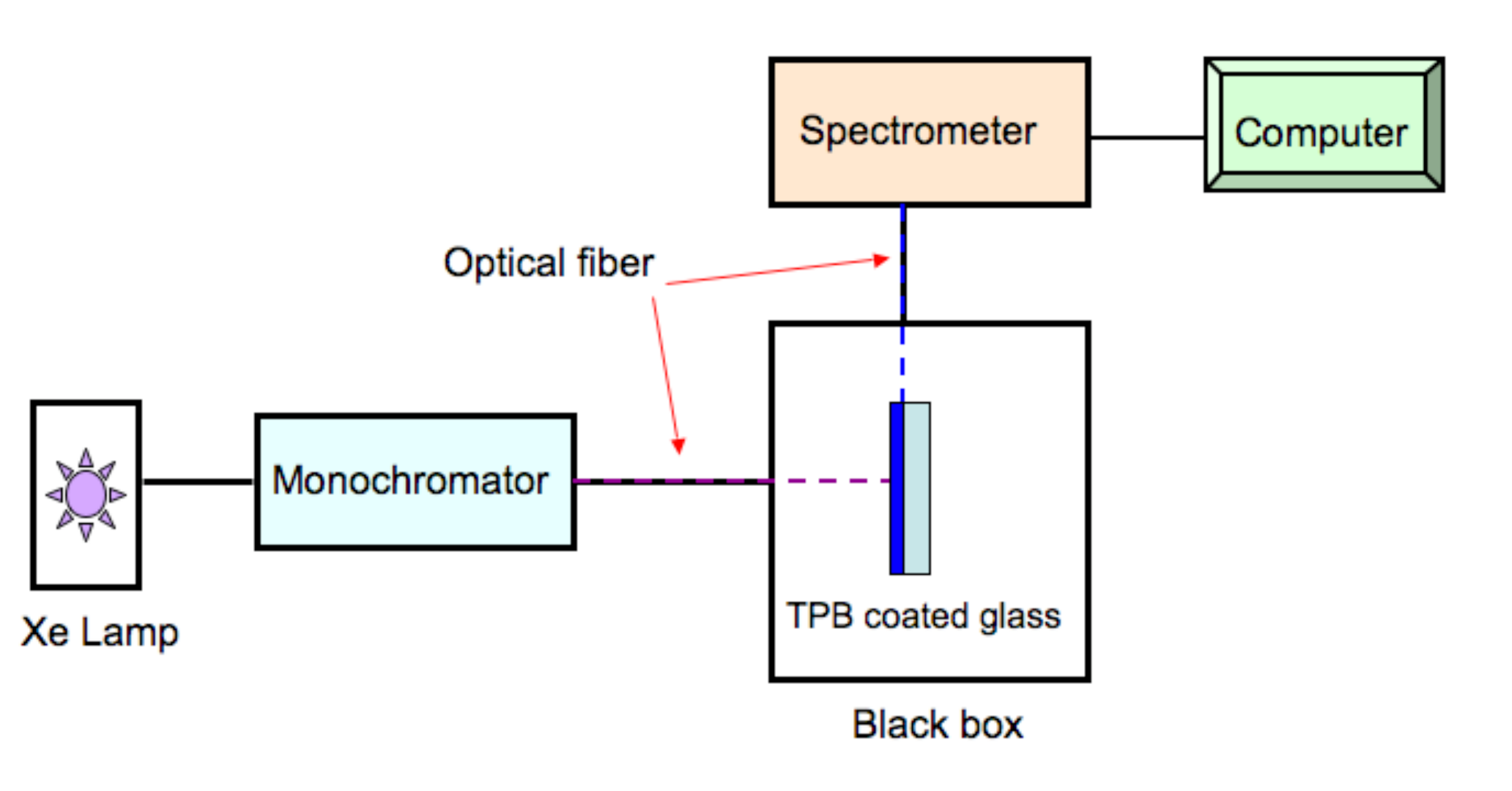}			
\caption{Diagram of the setup used to measure the TPB fluorescence using a coated glass plate.}
\label{fig:setup_diagram}
\end{figure}

\begin{figure}[h]
\centering
\includegraphics[width= 0.65\textwidth]{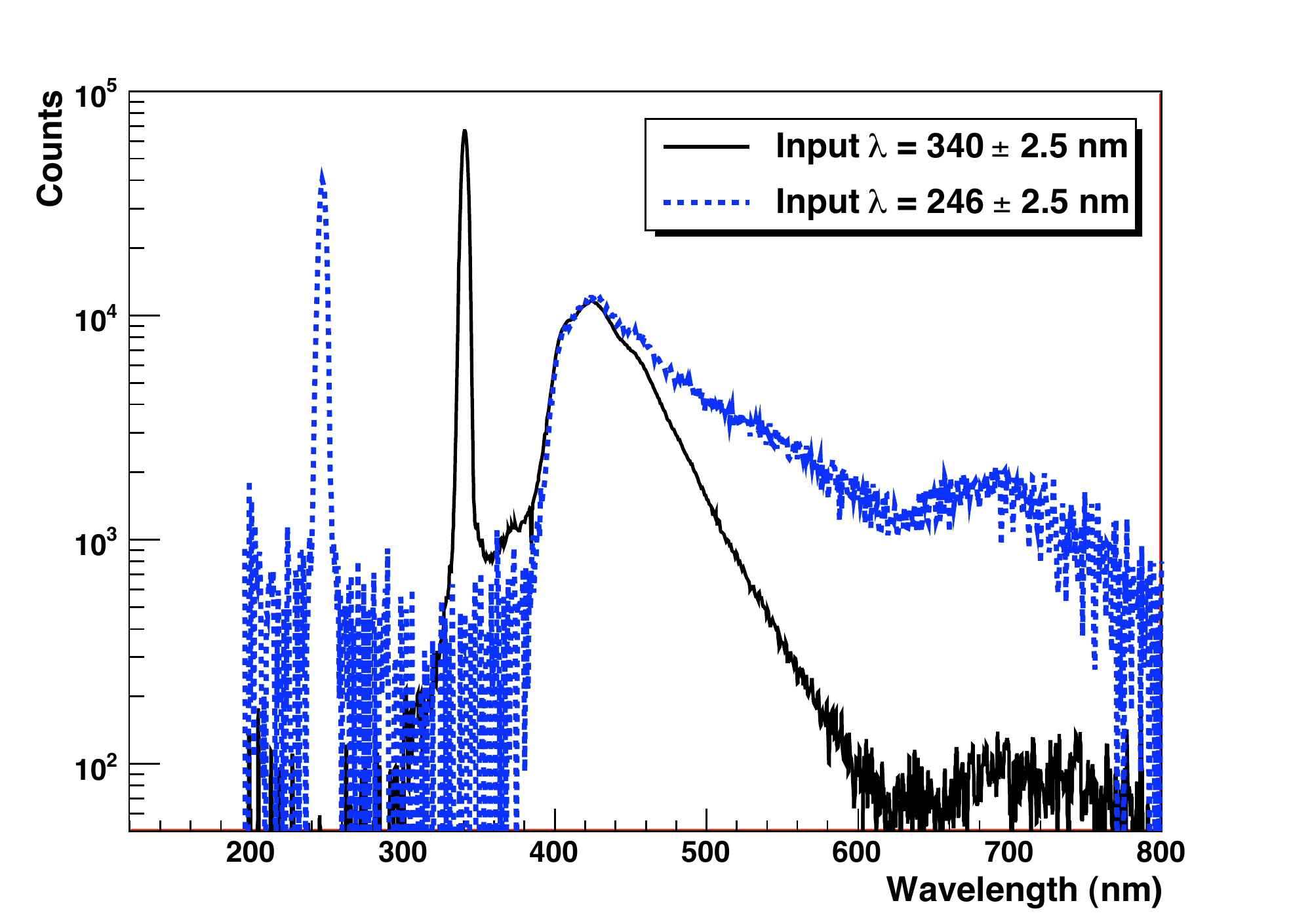}			
\caption{TPB emission spectra obtained from glass plate coated with  0.1~mg/cm$^2$ of TPB illuminated with 246~nm and 340~nm photons.}
\label{fig:TPB_spectrum}
\end{figure}

A glass plate of $30\times30$~mm$^2$, coated with 0.1~mg/cm$^2$ of TPB, was placed in a small black box to measure the fluorescence spectrum of the TPB (see figure~\ref{fig:setup_diagram}). A xenon lamp (Hamamatsu Photonics E7536, 150~W) coupled to a monochromator was used for the selection of the input wavelength. A spectrometer (Hamamatsu Photonics Multichannel Analyzer C10027) allowed to record the spectrogram of the output light from the TPB layer. The light from the monochromator to the black box was conducted through a quartz optical fiber, coupled to the box through an optical feedthrough. The spot of the input light covered an area of a few mm$^2$ of the glass surface.  The output light from the coated glass was collected by a lens, in the direction perpendicular to the input beam and conducted to the spectrometer by a quartz optical fiber. 
%%%
%%% 
The emission spectra of the TPB at the input wavelengths of $246\pm2.5$~nm and  $340\pm2.5$~nm were measured and are shown in figure~\ref{fig:TPB_spectrum}. 
%%%
%%% 

\begin{figure}[h]
\centering
\includegraphics[width=0.85\textwidth]{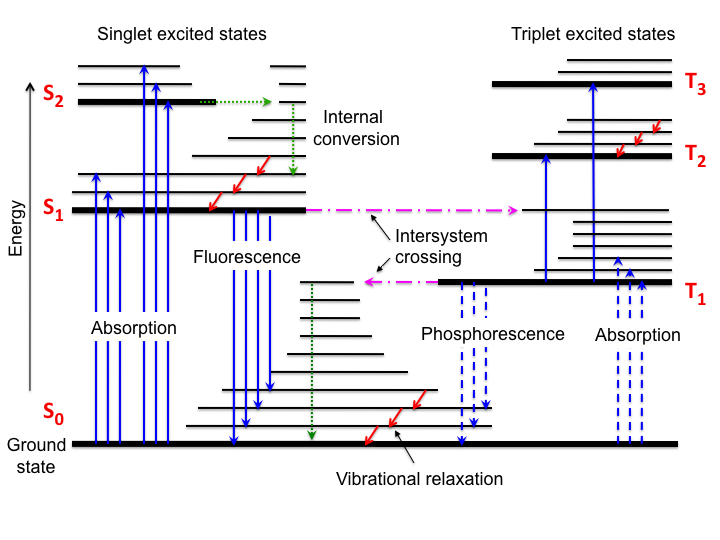}		
\caption{Typical diagram of the electronic energy levels of a molecule with singlet and triplet systems. The most important radiative (fluorescence and phosphorescence) and non-radiative (internal conversion, vibrational relaxation, intersystem crossing) transitions are shown (adapted from \cite{Haken}).}
\label{fig:energy_levels}
\end{figure}

As explained in the Molecular Physics literature \cite{Haken},\cite{Berlman}, TPB like some other organic compounds, fluoresces when its $\Pi$-orbital electrons (the $\Pi$-orbital is an electronic state in molecules, analogous to {\it p} electronic state in atoms) are excited either by ionizing particles or UV radiation.  
In figure~\ref{fig:energy_levels}, the typical scheme of the electronic energy levels of a molecule with singlet and triplet systems is shown. The fluorescence is a radiative transition from the vibrational levels (usually the lowest) of an excited electronic state (usually the first excited state S$_1$) into the vibrational levels of the ground state S$_0$ \cite{Haken}. Electrons excited to the vibrational states of S$_1$ relax within picoseconds to the lowest lying levels of 
S$_1$, before decaying to the ground state S$_0$, with the radiation of a photon at the compound's characteristic emission wavelength. 
The fluorescence decay time is typically in the nanosecond range. 
Besides fluorescence, an emission with a much longer decay time (typically in the microsecond range) called phosphorescence, is often observed, especially in the case of organic molecules. This denotes emission from an excited triplet state, i.e. from a state with a total spin quantum number S=1.  The longer decay time of phosphorescence is a result of the forbidden intersystem crossing for a transition from an excited triplet state into the singlet ground state  (see figure~\ref{fig:energy_levels}). In molecular physics, as in atomic physics, spin-orbit coupling allows forbidden singlet-triplet transitions to occur \cite{Haken}.

In the measurement of the TPB emission spectrum  shown in figure~\ref{fig:TPB_spectrum}, the peaks corresponding to non converted input light are seen, well separated from the fluorescence peak lying at $427\pm20$~nm. 
This fluorescence peak showed no dependance on the input wavelength in the UV range below 340~nm. It presents, moreover, a long tail at longer wavelengths which originates from the radiative decays of the S1 excited state to the multiple vibrational levels of the ground state.
In the TPB emission spectrum shown in figure~\ref{fig:TPB_spectrum}, another peak lying at  680~nm, most likely the phosphorescence peak,  was also observed. In the TPB compound, the UV light populates also the T1 triplet states which decay with the emission of longer wavelength photons.
These have however a small contribution to the TPB emission yield, as seen from the statistics shown in figure~\ref{fig:TPB_spectrum}.  

The TPB coated SiPMs of NEXT tracking plane, will therefore detect both the fluorescence and the phosphorescence from the emission of the TPB, exposed to the EL light. The much larger photon detection efficiency of the SiPMs at the fluorescence wavelength (up to 50\% at 430 nm compared to about 5\% at 680 nm ) makes this prompt component of the TPB emission widely dominate the light yield that will be detected by the tracking sensors. 

%%%
%%% 

\subsection{Deposition homogeneity}
\label{sec:homogeneity}

\begin{figure}[h]
\centering
\includegraphics[width= 7cm]{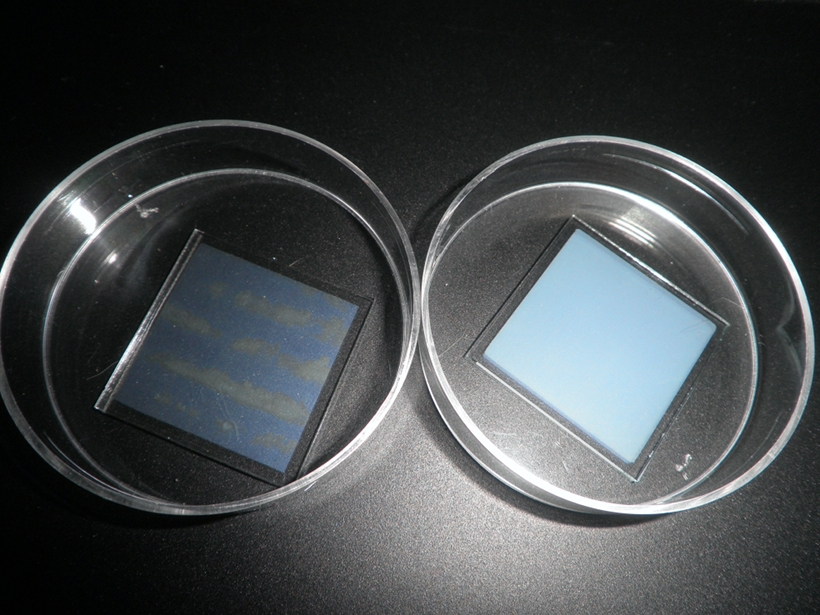}		
\caption{Photograph of glass samples coated without spinning (left) and with spinning (right)}
\label{fig:spinning}
\end{figure}

In the first coating trials, the TPB evaporation was performed without spinning the substrate. The homogeneity of the deposition was poor in these conditions as one can see in the glass sample on the left side of figure~\ref{fig:spinning}. The process was repeated with spinning at the maximum frequency of 50 rpm, keeping other parameters, as the deposition time and the vacuum level, unchanged. As a result, a quite homogenous deposition was formed, as seen on the right side of figure~\ref{fig:spinning}.
This homogeneity depends however on the exposition time and consequently on the deposition thickness, as the spinning velocity cannot be increased above its maximum value. This effect was studied using glass plates with identical dimensions (30$\times$30$\times$3~mm$^3$) coated with different TPB thicknesses (0.6, 0.2, 0.1 and 0.05~mg/cm$^2$). 

\begin{figure}[h]
\centering
\includegraphics[width=0.9\textwidth]{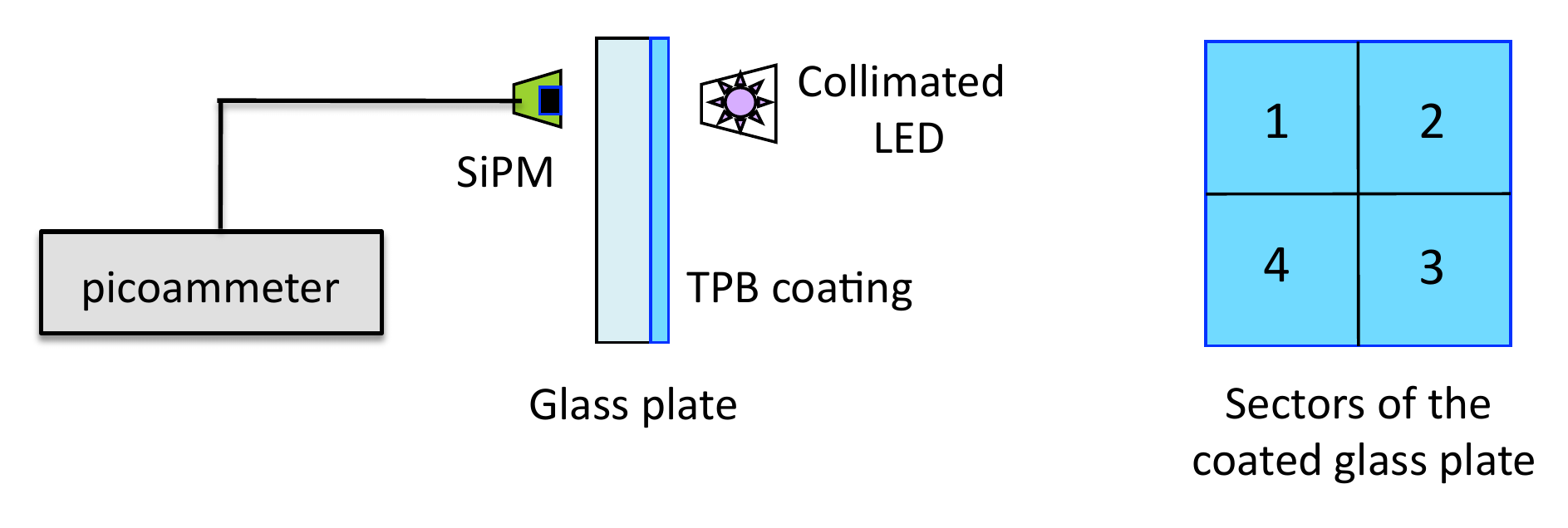}		
\caption{Scheme of the setup used for measuring the fluctuation of current in a SiPM induced by the converted light in different sectors of a coated glass plate of  30$\times$30~mm$^2$.} 
\label{fig:homogeneity_setup}
\end{figure}

\begin{figure}[h]
\centering	
\includegraphics[width= 0.65\textwidth]{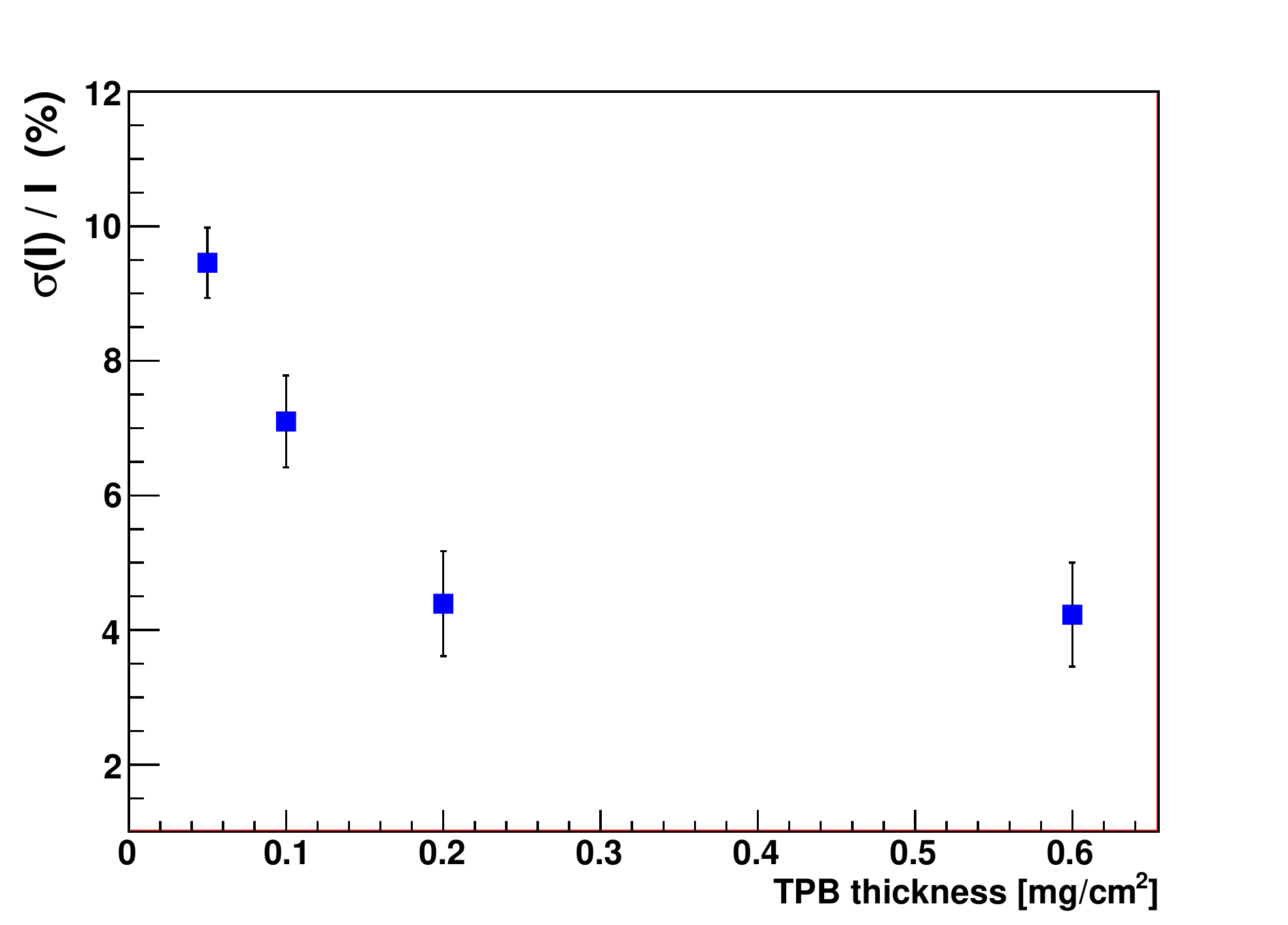}		
\caption{Relative standard deviation of current in a SiPM detecting light from different sectors of a coated glass plate as a function of TPB coating thickness.}
\label{fig:homogeneity_vs_thickness}
\end{figure}

It is known from the literature that the fluorescence yield of a TPB deposition depends on its thickness and is maximum for very thin coatings 
($< 0.15$~mg/cm$^2$) \cite{Lally}. Although this dependence is not linear, a qualitative indication of homogeneity in a TPB layer can thus be given by the surface dispersion of its fluorescence yield when illuminated by homogeneous UV light. We used for this measurement 
a LED with emission peak at 260~nm, placed in front of  the TPB coated side of the glass plate (see figure~\ref{fig:homogeneity_setup}). A non-coated SiPM (Hamamatsu MMPC S10362-11-025P) was placed on the other side, close to the surface, to measure the light converted successively in different coated sectors, as the glass plate was rotated.
The mean current in the SiPM for each light exposition was measured using a picoammeter (Keithley 6487) and the current relative standard deviation
$\sigma(I) / I_{mean}$ was drawn for each TPB coating considered. This standard deviation is represented in figure~\ref{fig:homogeneity_vs_thickness} as a function of thickness. The trend shown is compatible with that observed for the fluorescence yield as a function of thickness in reference \cite{Lally}. The relative standard deviation of the measured current was close to 10\% at the lowest thickness (0.05~mg/cm$^2$) and is slightly above 4\% for the thickest coating (0.6~mg/cm$^2$). These non-homogeneity levels which are relevant for coating millimeter scale photosensors for position measurements, may not be an issue for coating large surfaces as large PMT windows or internal walls of a detector vessel.     
%%%
%%% 
 %%%
%%% 
%%%
%%% 
\subsection{Transmittance}
\label{sec:transmittance}

The transmittance of the TPB at its emission wavelength was measured to evaluate the amount of converted light that would be reabsorbed  by the TPB layer as a function of its thickness. We used for this measurement
a LED emitting at $430\pm20$~nm as light source. Four glass plates of 30$\times$30~mm$^2$, coated respectively with 0.05, 0.1, 0.2 and 0.6~mg/cm$^2$  of TPB, were successively illuminated. The bare side of each glass plate was coupled to the window of  a 1 inch PMT (Hamamatsu R8520-406) whose anode current was measured with a picoammeter. A non-coated glass plate with the same dimensions of the coated plates was used as a reference.
The current in the PMT, induced by the light transmitted through the TPB layer, was compared to the current measured in the absence of TPB, using the reference glass and the same illumination conditions.
The transmittance of the reference glass at 430~nm was also measured using the same LED and PMT. This was $95.0 \pm 0.1$\%, which allows to evaluate
to about 5\%, the maximum loss of light due to reflections on the glass surface facing the LED.  In the TPB transmittance measurement,  we thus 
estimate the uncertainty of the reference current to be about 5\%, produced by the different light losses due to reflections on the glass-TPB and glass-air interfaces. 
This uncertainty was added to the statistical uncertainty, provided as the standard deviation from a sample of 100 current values, measured by the picoammeter at each light exposition.

\begin{figure}[h]   
\centering	
\includegraphics[width= 0.65\textwidth]{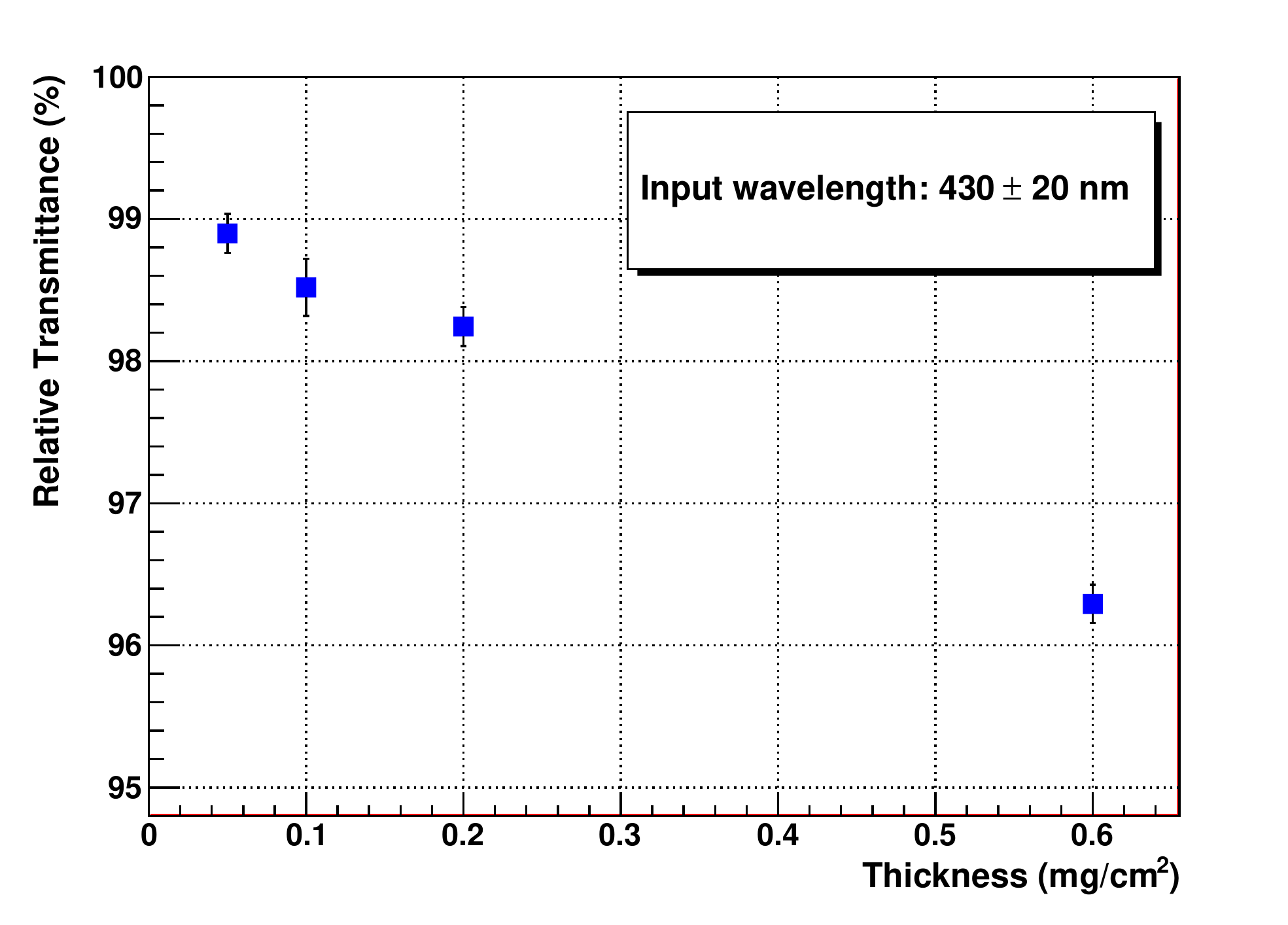}		
\caption{Transmittance of the TPB at its emission wavelength as a function of TPB thickness.}
\label{fig:self_absorption}
\end{figure}

The transmittance of the TPB at its fluorescence wavelength (430~nm) is shown in figure~\ref{fig:self_absorption} as a function of the thickness. The trend observed indicates an increase of the absorption with the TPB thickness. However, this amount of light absorbed, dissipated in the various non-radiative processes of the TPB molecule, remains below 4\% which is negligible for the position measurements.
TPB can thus be considered transparent to its fluorescence light in the range of thickness here considered ($\leq 0.6$~mg/cm$^2$).  
%%%
%%% 

\subsection{Ageing}
\label{sec:ageing}

For the very first deposition trials, the TPB powder was taken from a sealed bottle as provided by the manufacturer and stored at ambient conditions
in the laboratory. The depositions produced with this TPB were repeatedly non-uniform despite of optimal temperature, vacuum and spinning conditions in the evaporator. The depositions were also difficult to characterize using the profilometer. These results were a clear indication of degradation of the molecule by the environmental conditions of storage, mainly temperature ($\approx 22$ C$^{\circ}$) and sunlight.
Indeed, successful evaporations from a newly purchased TPB powder, stored in dark at low temperarure (2 - 8~C$^{\circ}$), confirmed this assumption. 

The long term stability of the TPB depositions is an important issue for NEXT. As reported in previous studies \cite{Lally},\cite{Jerry}, TPB degrades in air because of the oxidation, hydration and exposure to sunlight. The coated SiPMs of the NEXT-DEMO tracking plane were stored in vacuum and were re-tested after long storage time to investigate possible variations in their response due to ageing of TPB. The results of this measurements
are presented in section~\ref{sec:DB_ageing}. 

%%%%%%%%%%%%%%%%%%%%%%%%%%%       section 5         %%%%%%%%%%%%%%%%%%%%%%%%%%%%%%%%%%%%%%%%%%%%%%%%%%%%%%
\section{Response of coated SiPMs}
\label{response_sipms}

\subsection{Dependence on coating thickness}

An important issue when coating the SiPMs for NEXT is the determination of the coating thickness which fulfills the requisites of optimal conversion efficiency and deposition homogeneity.  
The TPB conversion efficiency is reported in the literature to be optimal at  deposition thicknesses 
between 0.05 and 0.15~mg/cm$^2$ \cite{Lally}. For large PMT windows, the TPB coating thickness of 0.05~mg/cm$^2$ is found to provide the optimal PMT response \cite{Boccone},\cite{Benetti}.\\
However, coating SiPMs of 1~mm$^2$ active area with less than 0.2~mg/cm$^2$ of TPB would significantly deteriorate the response uniformity of the SiPMs for the position measurement as shown in figure~\ref{fig:homogeneity_vs_thickness}. 
Coating thicknesses less than 0.1~mg/cm$^2$ were therefore discarded and the response of SiPMs coated respectively with 0.1 and 0.2~mg/cm$^2$ of TPB were compared to assess the choice of the TPB thickness for the NEXT tracking plane.
 
Two SiPM boards (B1) and (B2) (made out of FR-4 PCB) containing each 5 Hamamatsu S10362-11-050P MPPCs, were respectively coated with 0.1 and 0.2~mg/cm$^2$ of TPB.
The SiPMs were biased individually using the operation voltages provided by the manufacturer. This ensured a uniform response of the sensors within the board. The SiPMs were illuminated successively by a collimated LED emitting at 260~nm, operated in continuous mode. Their output current was measured with a picoammeter, prior to and after coating, at the same illumination and temperature conditions. 

\begin{figure}[h]
\centering
\includegraphics[width= 0.65\textwidth]{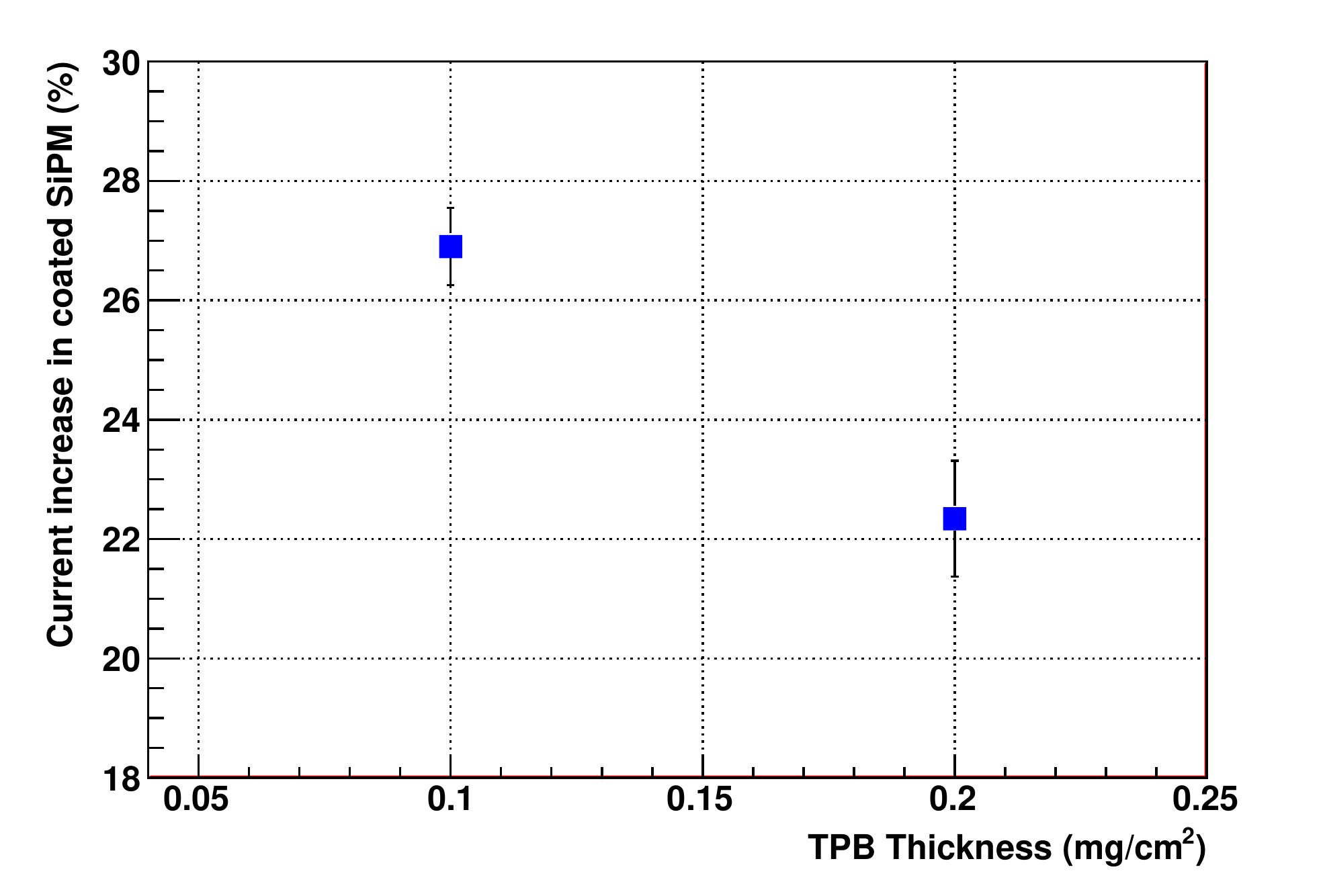}		
\caption{Response of the coated SiPMs to UV light (260 nm) as a function of coating thickness. For each thickness, the average of the currents measured in each of the 5 SiPMs of a board was taken.}
\label{fig:fluorescence_vs_thickness}
\end{figure}

The average current in board B1 and B2, prior to coating, were $47.2\pm 4.3$~$\mu$A and $60.0\pm 2.8$~$\mu$A respectively, after subtraction of the dark current (typically less than 0.5~$\mu$A).  
To evaluate the response of the coated SiPMs as a function of the two TPB thicknesses, the average relative increase of current due to TPB was taken and represented in figure~\ref{fig:fluorescence_vs_thickness}. 
In this figure the SiPMs with 0.2~mg/cm$^2$ of TPB  have about 19\% less response than the SiPMs with the half TPB thickness. This decrease of converted photon yield with thickness confirms the trend reported in the literature.
However, it does not provide the absolute difference in conversion efficiency between both thicknesses but a lower limit, as the measurement is  affected by the high illumination level used. Indeed, much higher response of the coated SiPMs was also measured at similar wavelength and lower illumination levels in further measurements (see figure~\ref{fig:Xenon_lamptest}).  

 The choice of the coating thickness for NEXT-DEMO SiPMs was thus decided for 0.1~mg/cm$^2$  as it provides a good compromise between 
high enough fluorescence efficiency and good enough deposition homogeneity.
%%%
%%% 

\subsection{Dependence on wavelength}

\begin{figure}[h]
\centering
\includegraphics[width= 6.5cm]{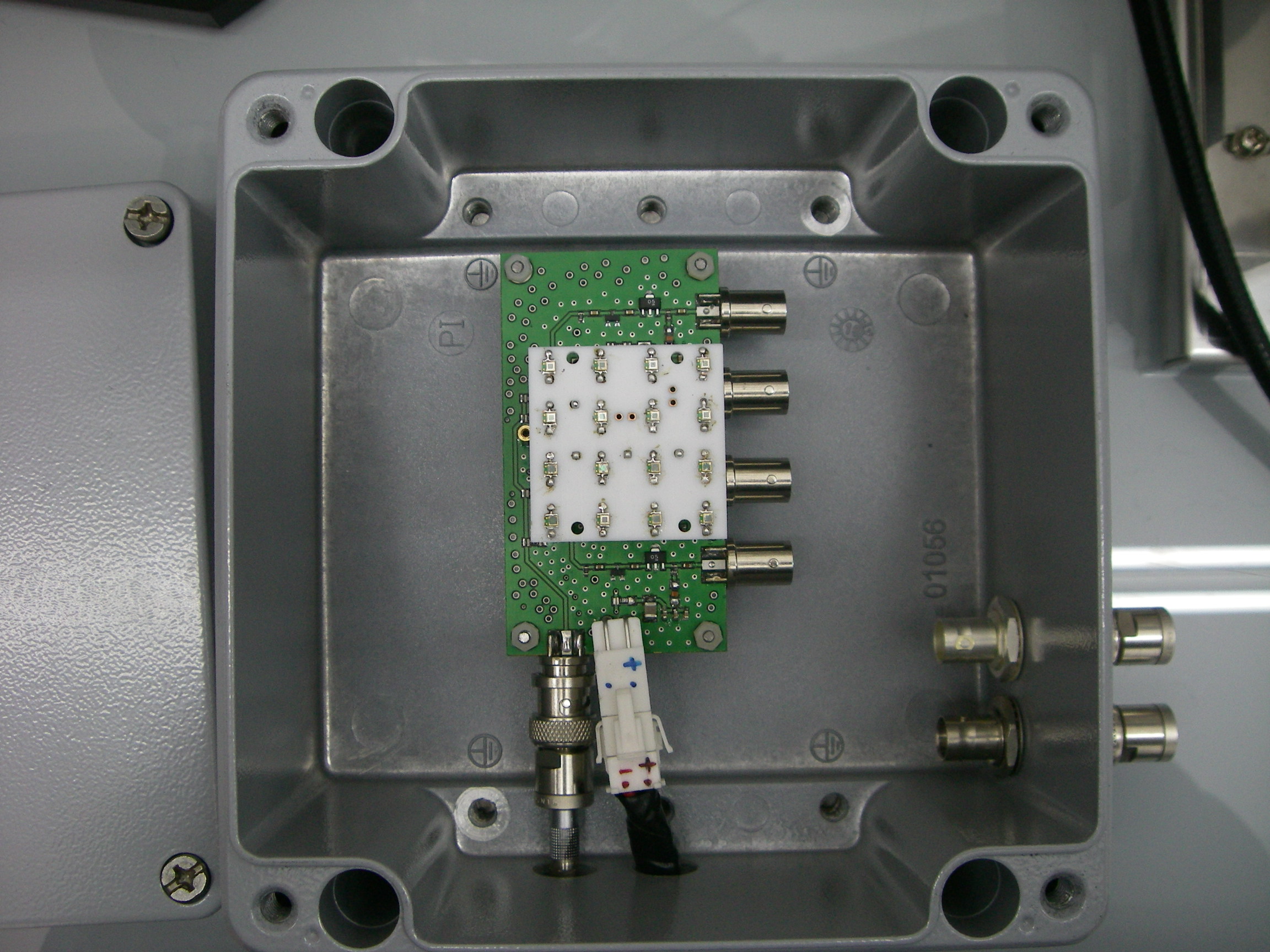}
\includegraphics[width= 6.5cm]{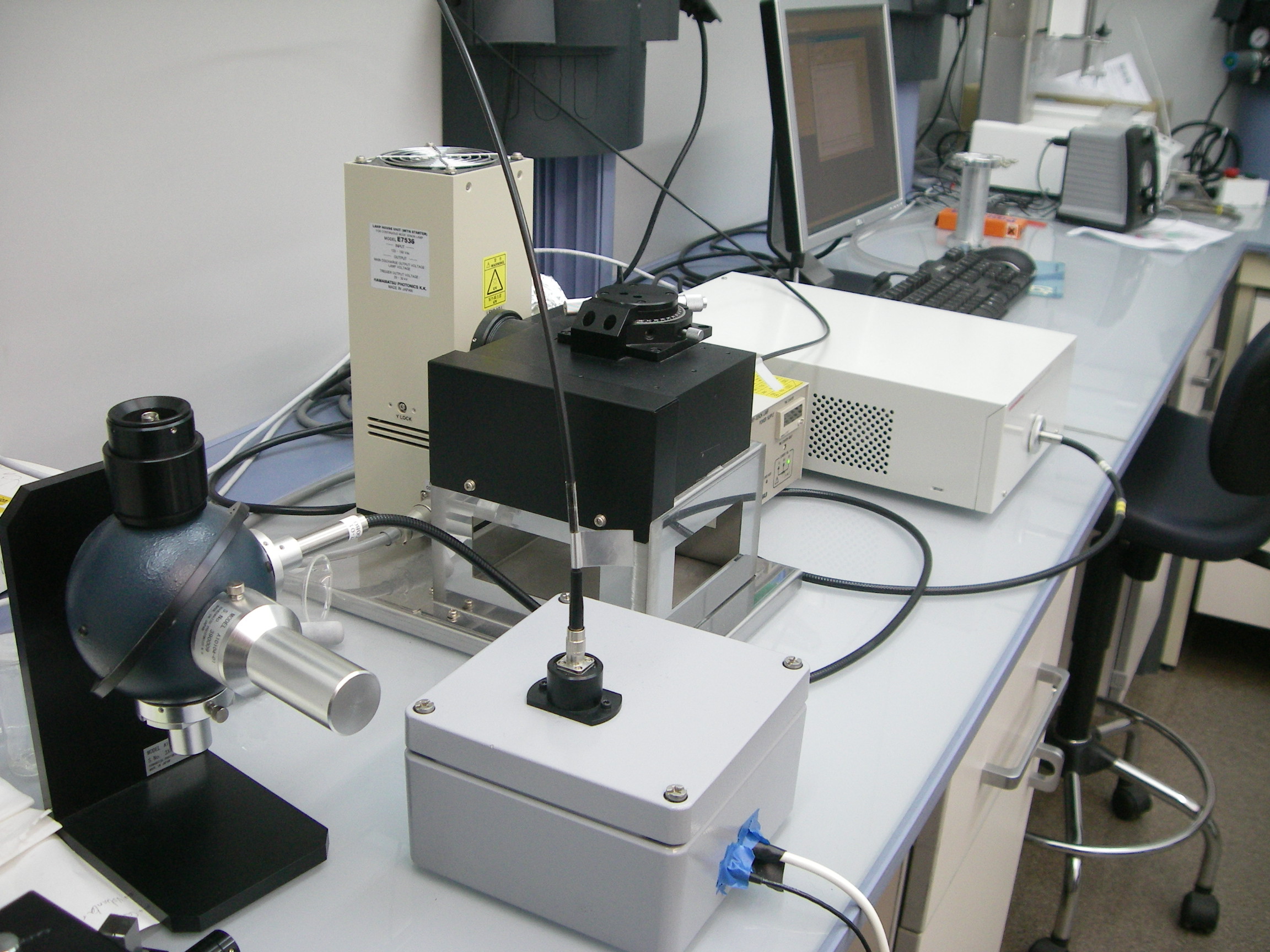}		
\caption{(left) The light-tight box with one coated SiPM DB plugged on a readout PCB.  
(right) The closed box with the optical fiber from the monochromator output plugged on top. }
\label{fig:Test_bbox}
\end{figure}

\begin{figure}[h]
\centering
\includegraphics[width= 0.65\textwidth]{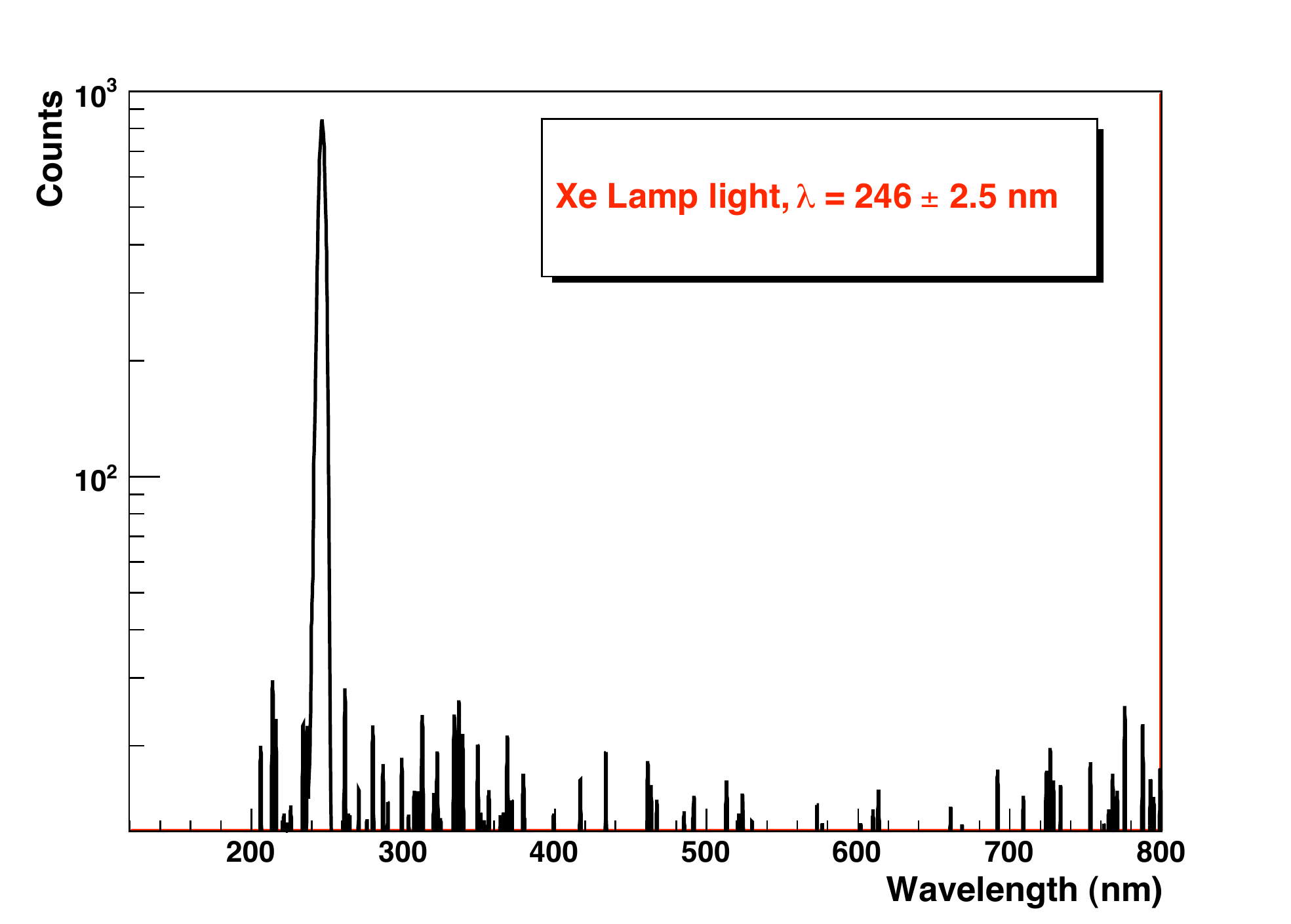}	
\caption{Emission spectrum of the xenon lamp after the monochromator with filtering set to 246~nm. }
\label{fig:monochromator_spectrum}
\end{figure}

The 18 DBs of the NEXT-DEMO tracking plane were coated with 0.1 mg/cm$^2$ of TPB and then stored in a vacuum chamber before mounting on the mother-board for operation in a N$_2$ chamber or inside the Xe TPC. 
The response of the SiPMs of the coated DBs as a function of wavelength was measured using a xenon lamp coupled to a monochromator.  This allowed to select the input wavelength down to $246$~nm.  The DB was plugged onto a readout PCB placed inside a light-tight box as shown in figure~\ref{fig:Test_bbox}.
The optical fiber from the monochromator was plugged onto the top cover of the box through an optical feedthrough. A diffraction lens coupled to the optical feedthrough allows to illuminate the entire SiPM board placed in the bottom of the box.
%%%
%%%     
The spectrum from the monochromator used to illuminate the SiPMs was analyzed with a spectrometer. An example is shown in 
figure~\ref{fig:monochromator_spectrum} for the monochromator output set to 246~nm. The average current from the 16 coated SiPMs of a DB was measured, for each input wavelength, using an electrometer and compared to the average current from a non-coated DB, exposed to the same illumination conditions. 
The average dark currents measured in the coated and non-coated DBs were $8.0\pm0.1$~nA and  $9.6\pm0.1$~nA respectively.    

\begin{figure}[h]
\centering
\includegraphics[width= 0.65\textwidth]{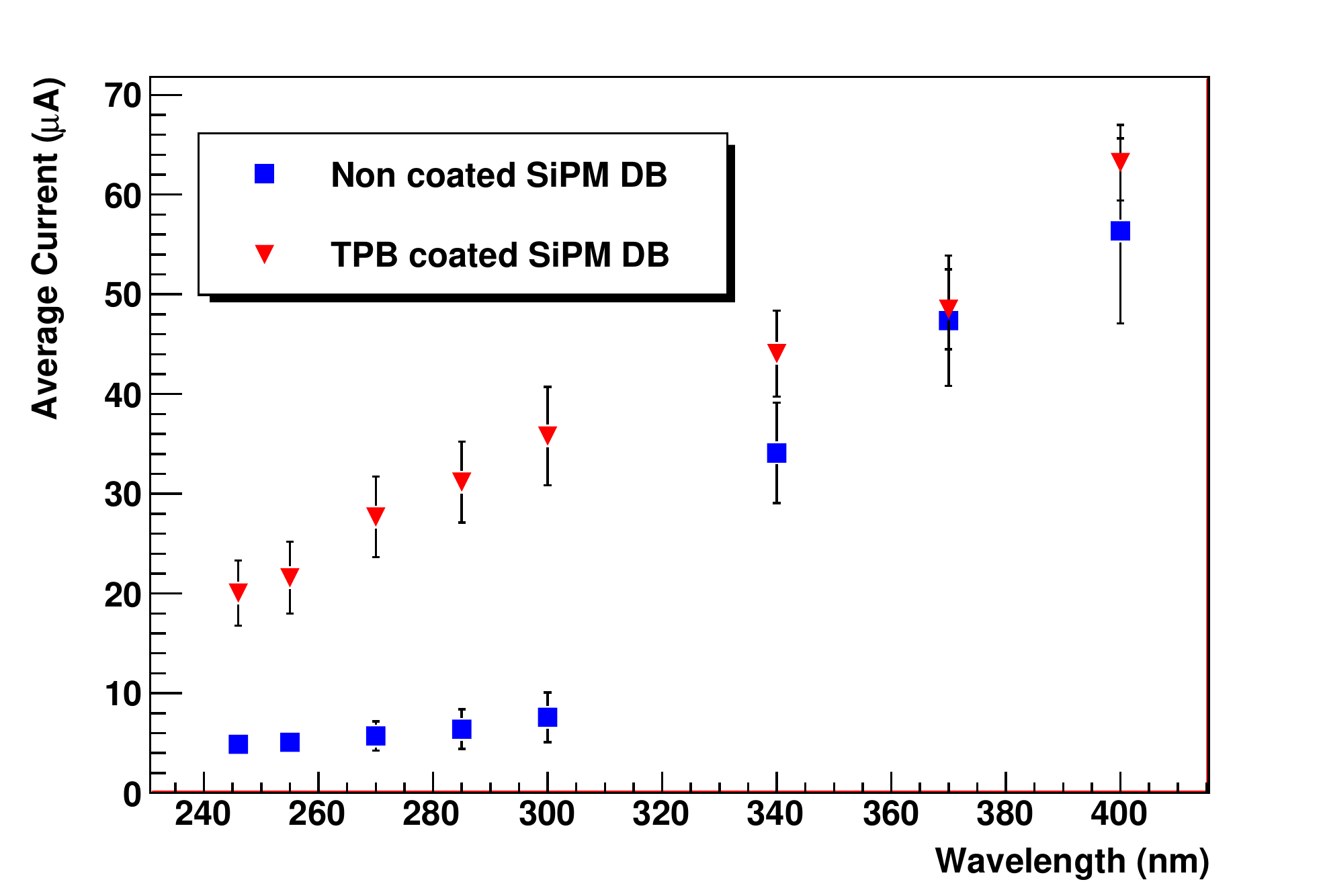}			
\caption{Average current in the 16 SiPMs of a coated DB as a function of wavelength, compared to the average current of a non coated SiPM DB.
The SiPMs are illuminated at different wavelengths using a xenon lamp coupled to a monochromator. The decrease in the SiPM average current at short wavelengths reflects the decrease of the UV light yield of the xenon lamp as the wavelength decreases.}
\label{fig:Xenon_lamptest}
\end{figure}

In figure~\ref{fig:Xenon_lamptest}, the average current measured after dark current subtraction, is plotted as a function of the input wavelength for both coated and non-coated SiPMs. As it is seen, the sensitivity of the non-coated SiPMs decreases significantly at wavelengths below 340~nm, although it is still 
significant at 246~nm. Indeed, the average current of the non-coated SiPMs induced by the intense Xe lamp is close to 5~$\mu$A, significantly above the dark current. On the other hand, a significant increase in the response of the coated SiPMs was observed in the UV range below 340~nm and down to 246 nm. In this spectral range, the SiPMs response is enhanced by a factor 4 to 5 due to TPB coating and appears to decrease with the short wavelengths.  
This effect due to the decrease of the light yield of the xenon lamp at short wavelengths, does not reflect a decrease in the TPB conversion efficiency. Indeed, the TPB conversion efficiency does not depend on the input wavelength in the UV range as shown in reference \cite{Lally} and discussed in section~\ref{sec:fluorescence} of this paper.  \\
In the visible region of the spectrum ($> 360$~nm) however, the coated and non-coated SiPMs have a similar response within measurement uncertainty.  

 %%%
%%%      
%%%
%%%          

\begin{figure}[h]
\centering
\includegraphics[width= 0.7\textwidth]{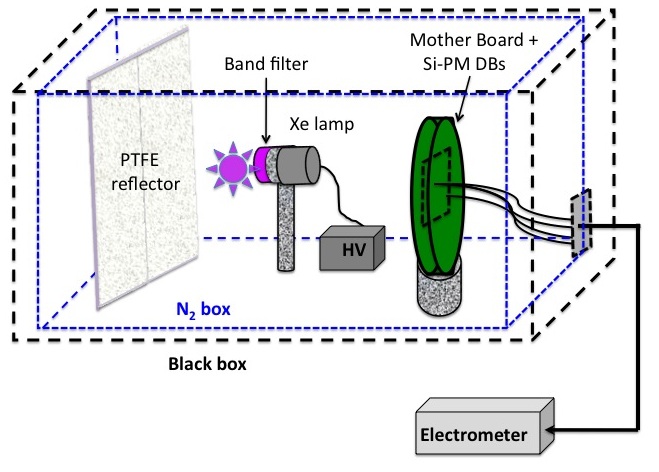}	
\caption{Scheme of the experimental setup used for testing the VUV sensitivity of the SiPMs coated and non-coated with TPB.}
\label{fig:N2_box_setup}
\end{figure}

The response of the TPB-coated SiPMs in the VUV spectrum range of interest for NEXT was investigated in our laboratory using a Xe flash lamp (Hamamatsu L2358, 15 W, synthetic silica window and 100 Hz maximum repetition rate) coupled to a band filter (eSource Optics \cite{bandfilter}) with 
band spectrum peaked at 173.0$\pm 20.0$~nm. The lamp with its band-filter was directed towards a PTFE reflector plane parallel to the mother board and placed at 40 cm distance from it as shown in the schematic of figure~\ref{fig:N2_box_setup}. This arrangement ensures a quite uniform illumination of the daughter boards plugged onto the mother board. The whole setup was placed inside a glove-box filled with N$_2$ to ensure VUV light transmission without absorption by the oxygen in the air. The N$_2$ box was in turn enclosed in a light-tight box. 

\begin{figure}[h]
\centering
\includegraphics[width= 0.58\textwidth]{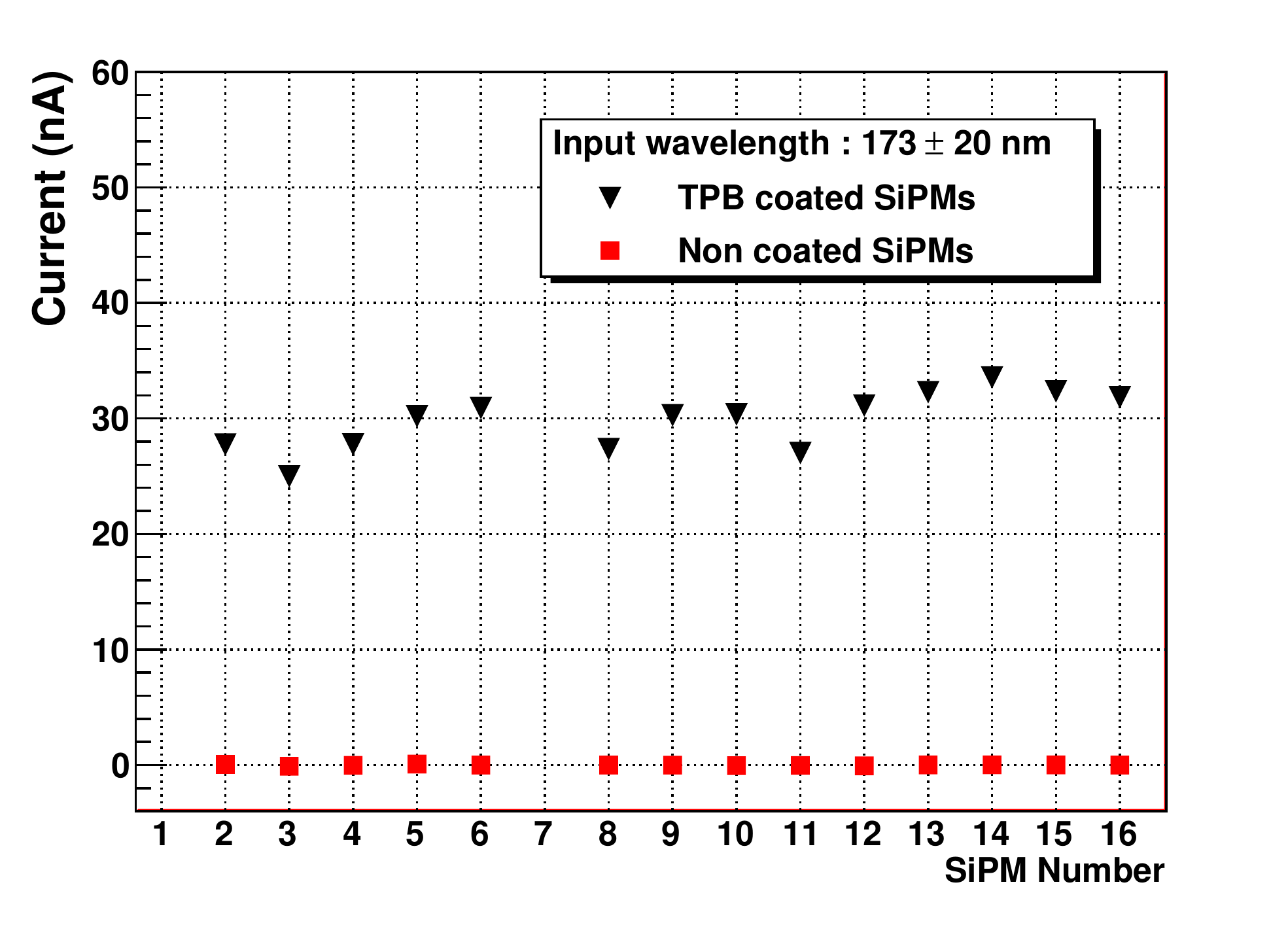}
\includegraphics[width= 0.4\textwidth]{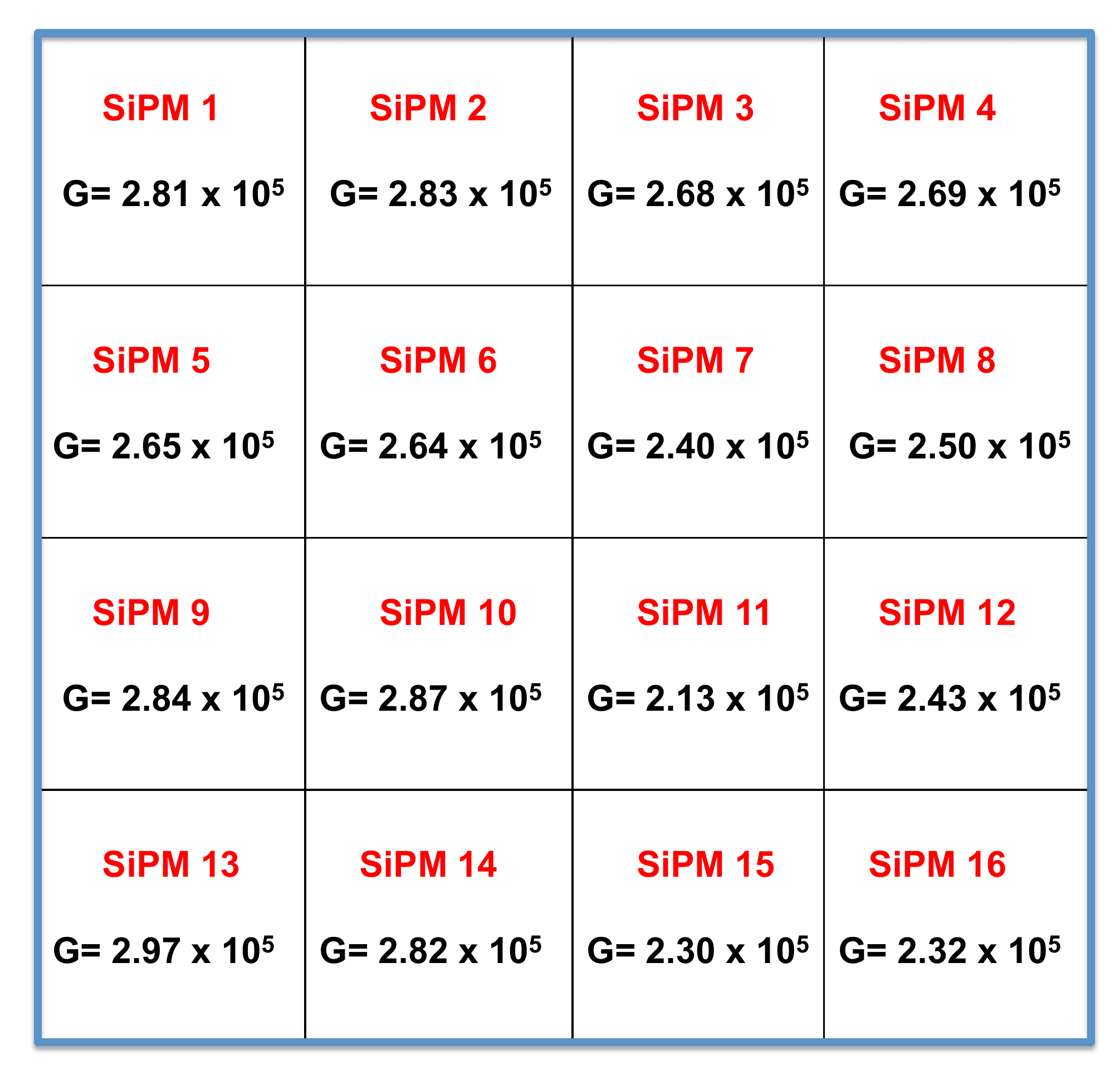} 	
\caption{(left) Response to homogeneous VUV light of a TPB-coated DB and a non-coated DB. (right) Configuration of the SiPMs in the daughter boards and their respective gain. }
\label{fig:currents_173nm}
\end{figure}

\begin{figure}[h]
\centering
\includegraphics[width= 0.65\textwidth]{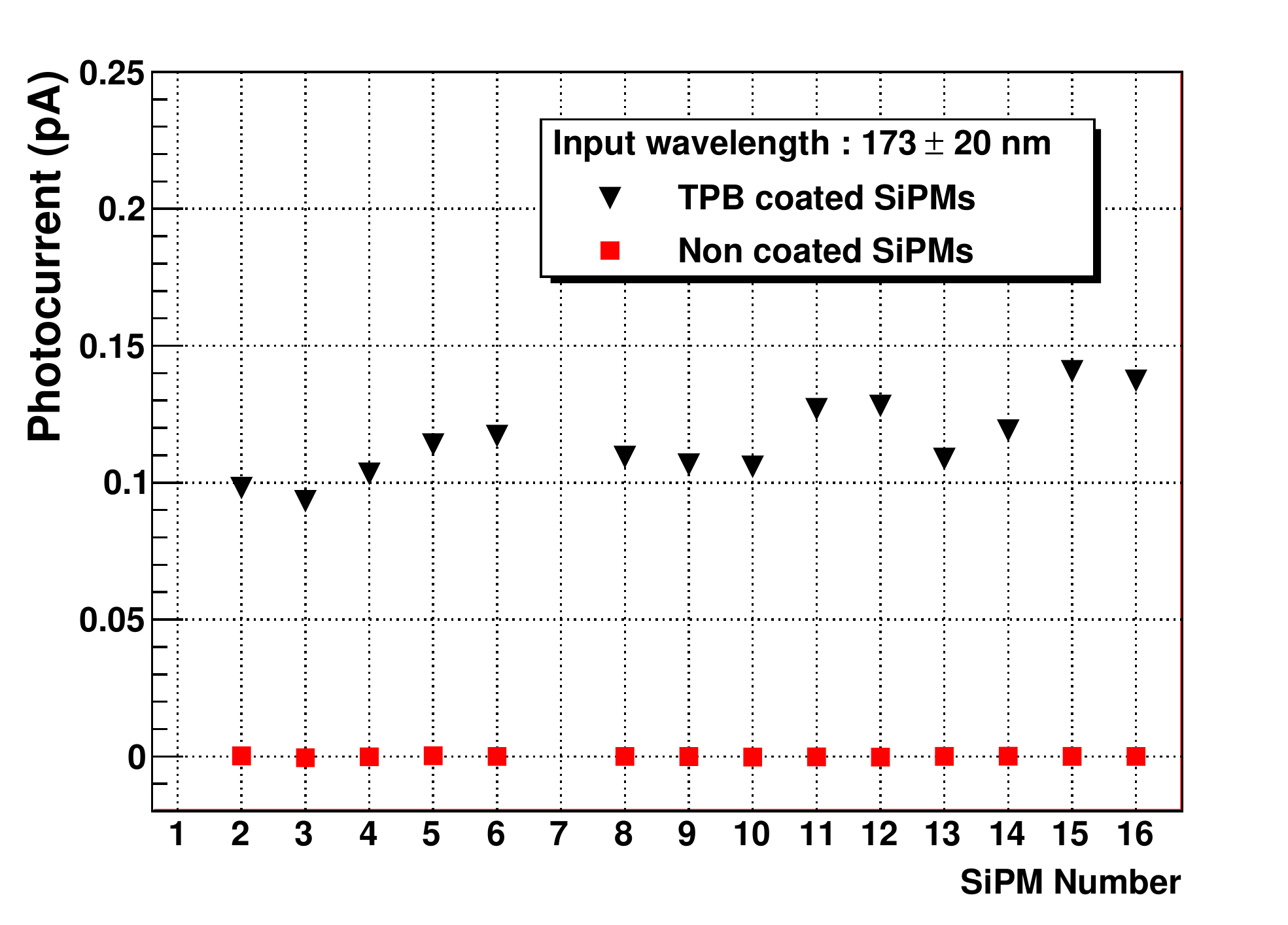} 
\caption{SiPMs photocurrent induced by the homogeneous VUV light of a TPB-coated DB and a non-coated DB.}
\label{fig:photocurrents_173nm}
\end{figure}

Two sample DBs with 14 SiPMs were used, both with 9\% gain dispersion, measured using the single photon response of the SiPMs \cite{Alvarez} prior to coating. One of these DBs was coated with 0.1 mg/cm$^2$ of TPB. The current from the individual SiPMs was measured using an electrometer (Keithley 6517B) which also supplied the SiPM bias.  
%%%
%%%  
In figure~\ref{fig:currents_173nm}, the response of the coated and non-coated SiPMs at the xenon scintillation wavelength is shown. The dark currents (typically 8 nA) were first measured and subtracted from the currents induced by the VUV light.  As it can be seen in the figure, the non-coated SiPMs do not respond to the input light at 173 nm, whereas the TPB-coated SiPMs have a significant and quite uniform response at this short wavelength. The current dispersion of the coated DBs was indeed 8\%, including dispersion of the SiPM gain and other contributions as the residual non-uniformity in the incident light and in the TPB coating.  In figure~\ref{fig:photocurrents_173nm} the SiPMs photocurrent (current normalized to the gain) is represented to evaluate these latter effects. 
%%%
%%% 
The photocurrent dispersion of the coated DB was found of 11~\%, which includes the non-uniformity in the TPB coating and in the incident light and the dispersion in the PDE of the SiPMs. Taking into account these two latter uncorrelated (and unknown) factors, we consider the TPB deposition of this DB sample homogeneous enough for tracking. This result could be even better in other coated DBs, stored in vacuum and not extensively used for laboratory tests.
%%%
%%% 

\subsection{Response after long-term storage}
\label{sec:DB_ageing}
\begin{figure}[h]
\centering
\includegraphics[width=0.65\textwidth]{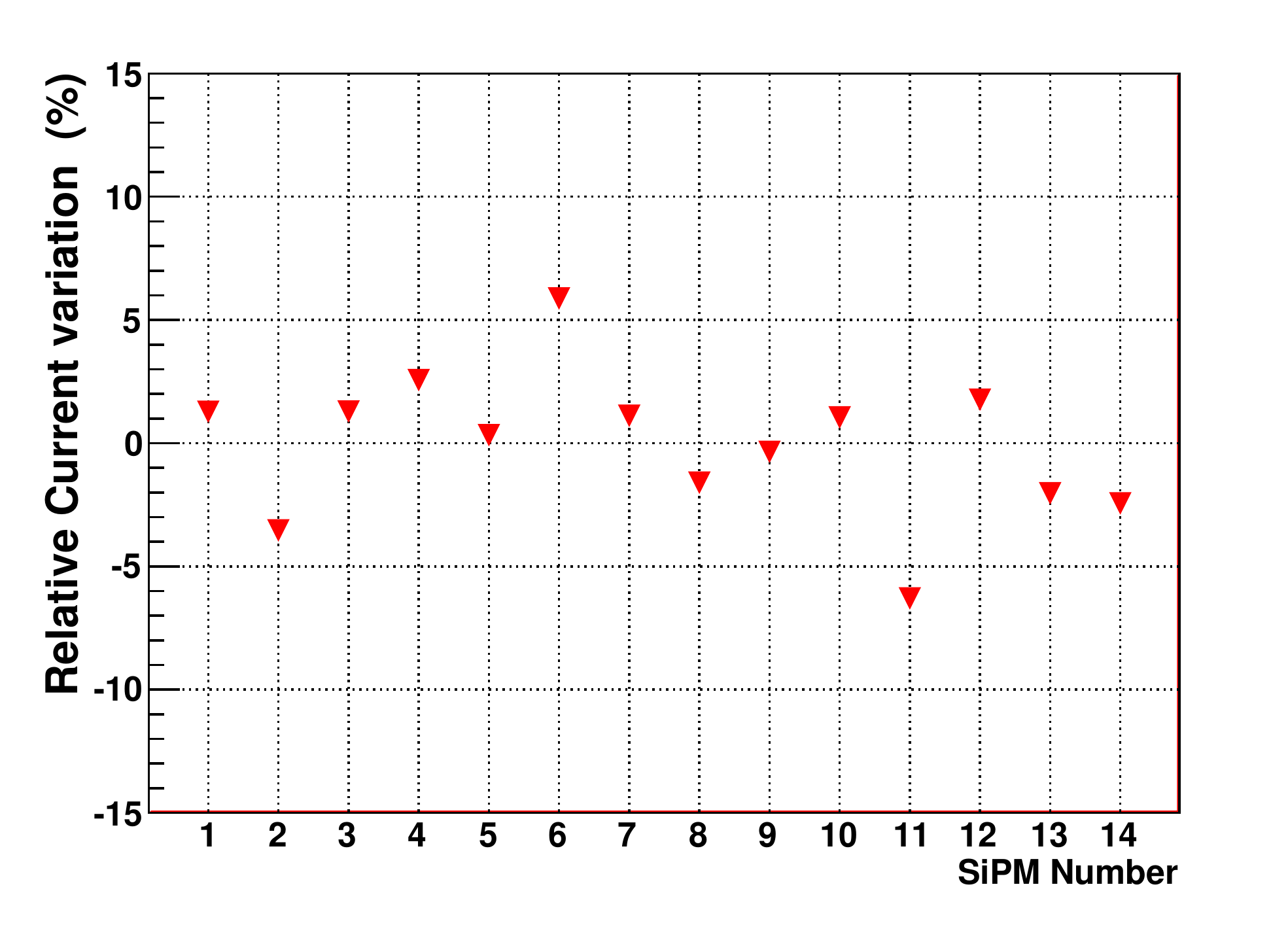}	
\caption{Relative current variation of 14 coated SiPMs, illuminated with a LED emitting at 240~nm, after 9 months of storage in vacuum}
\label{fig:ageing_sipms}
\end{figure}

 As mentioned before, the coated SiPM DBs of the NEXT-DEMO tracking plane were stored in a vacuum chamber at pressure $< 1$~mbar. A sample of these DBs was re-tested after 9 months of storage using a LED emitting at 240~nm, operated in pulsed mode (pulse width=30~ns, frequency=1~kHz), and illuminating individually each of the 16 SiPMs of the coated DB. The current measured  (with dark current subtracted) after storage was compared to that measured immediately after the coating process. The relative current variation of the 14 SiPMs tested is shown in 
figure~\ref{fig:ageing_sipms}. The average of all currents in the SiPMs has less than 1\% relative variation, which is well within the experimental uncertainties.
Indeed, considering the possible variation of LED intensity and ambient temperature after 9 months, the current variation of the coated SiPMs observed indicates no evidence of ageing effects in the TPB coatings stored in a moderate vacuum.
%%%
%%%   
Additional measurements with coated DBs stored in high pressure xenon are foreseen and will be reported in a further publication.   
  
\section{Conclusions}
The reconstruction of the tracks and topology of the events in the NEXT gaseous Xe TPC is a key issue for background rejection in the double-beta decay search. A prototype of the NEXT tracking system composed of 248 MPPCs (Hamamatsu S10362-11-025P) was built. These sensors offer important advantages for tracking purposes over other photosensors but they are not sensitive to the Xe scintillation. A protocol for coating the SiPMs with TPB used as WLS, was developed with particular precautions in obtaining clean and uniform coatings, with optimal fluorescence efficiency. The characterization of the coated samples with different UV light sources shows high quality TPB coatings on the SiPMs of the NEXT-DEMO tracking system. Furthermore, the coated SiPMs show a significant and uniform response to the Xe scintillation wavelength 
($\approx175$ nm) compared to the non-coated ones, which are completely non-sensitive to this wavelength. 
The response of coated SiPMs, after 9 months of storage in a moderate vacuum (< 1~mbar), shows no evidence of ageing effects in the coatings.  NEXT-DEMO tracking system is presently tested in a N$_2$ glove-box with a Xe lamp and will be commissioned for operation inside the TPC in the near future.

\acknowledgments
We acknowledge the Spanish MICINN for the Consolider Ingenio grants under contracts CSD2008-00037, CSD2007-00042 and CSD2007-00010 and for the research grants under contract FPA2008-03456 and FPA2009-13697-C04-01 part of which come from FEDER funds.\\
The Portuguese team acknowledges support from FCT and FEDER through program COMPETE, project PTDC/FIS/103860/2008.\\
J. Renner acknowledges the support of the U.S. Department of Energy Stewardship Science Graduate Fellowship, grant number DE-FC52-08NA28752.


\begin{thebibliography}{9}

\bibitem{CDR} NEXT collaboration: V. Álvarez et al., \emph{The NEXT-100 experiment for neutrinoless double beta decay searches (Conceptual Design Report).}  {\bf arXiv:1106.3630}

\bibitem{LBNL} Azriel Goldschmit et al., \emph{High-pressure xenon gas TPC for neutrino-less double-beta decay in $^{136}Xe$: Progress towards the goal of 1\% FWHM energy resolution}. \emph{Submitted to the 2011 IEEE Nuclear Science Symposium.}

\bibitem{Yahlali} Nadia Yahlali, Igor G. Irastorza, \emph{First Next prototypes for double-beta decay search}, { \emph{Nucl. \ Instrum. \  Meth.} {\bf A 581}~(2011)~162}.

\bibitem{Lally} C.H.~Lally, G.J. Davies, W.G. Jones, N.J.T. Smith, \emph{UV quantum efficiencies of organic fluors}, { \emph{Nucl. \ Instrum. \  Meth.} {\bf B 117}~(1996)~421}.

\bibitem{Gehman} V.M. Gehman et al., \emph{Fluorescence efficiency and visible re-emission spectrum of tetraphenyl butadiene films at extreme ultraviolet wavelengths}, { \emph{Nucl. \ Instrum. \  Meth.} {\bf A 654}~(2011)~116}.

\bibitem{Boccone} V.~Boccone et al., \emph{Development of wavelength shifter coated reflectors for the ArDM argon dark matter detector}, \jinst{4}{2009}{P06001}. 

\bibitem{Benetti} P.~Benetti, C. Montanari, G.L. Raseli, M. Rossella, C. Vignoli, \emph{Detection of the VUV liquid argon scintillation light by means of glass-window photomultiplier tubes}, \emph{Nucl. \ Instrum. \  Meth.} {\bf A 505}~(2003)~89. 

\bibitem{Hamamatsu} http://jp.hamamatsu.com/products/division/ssd/

\bibitem{Polyflon} http://www.polyflon.com

\bibitem{Alvarez} V.~\'Alvarez et al., \emph{Design and characterization of the SiPM tracking system of NEXT1-IFIC}, JINST preprint. 

\bibitem{Herrero} V. Herrero et al., \emph{Readout electronics for the SiPM tracking plane in the NEXT-1 prototype}, 
in Proceedings of the 6th International Conference on New Developments  in Photodetection (NDIP2011), to be published in { \emph{Nucl. \ Instrum. \  Meth.} {\bf A}}.

\bibitem{Renker} D. Renker and E. Lorenz, \emph{Advances in solid state photon detectors}, \jinst{4}{2009}{P04004}.

\bibitem{Aldrich} Aldrich Chemical Company Inc., http://www.sigmaaldrich.com.

\bibitem{Burton}W.M.~Burton and B.A.~Powell, \emph{Fluorescence of Tetraphenyl-Butadiene in the Vacuum Ultraviolet}, {\emph{Appl. Optics} {\bf 12}~(1973)~87}.

\bibitem{QCM} http://www.inficonvacuumcoating.com/en/index.html.

\bibitem{Sauerbrey} G. Sauerbrey, \emph{Verwendung von Schwingquarzen zur w\"agung d\"unner Schichten und Microw\"agung}, {\emph{Z.  Phys.} {\bf 155}~(1959)~206-222.}

\bibitem{Ambios} http://www.ambiostech.com.

\bibitem{Haken} H. Haken, H.C. Wolf, \emph{Molecular Physics and Elements of Quantum Chemistry}. Second Edition, Springer.  

\bibitem{Berlman} I.B.~Berlman, \emph{Handbook of fluorescence spectra of Aromatic Molecules}, (Academic Press, New York), 1971.

\bibitem{Jerry} R. Jerry, L. Winslow, L. Bugel and J.M. Conrad, \emph{A study of the Fluorescence Response of Tetraphenyl-butadiene}, {\bf arXiv:1001.4214}. 

\bibitem{bandfilter} http://www.esourceoptics.com/

\end{thebibliography}
\end{document}